\documentclass[twocolumn]{aastex631}

\newcommand{\teff}{$T_{\rm eff}$}
\newcommand{\logg}{$\log{g}$}
\newcommand{\feh}{[Fe/H]}
\newcommand{\ALi}{$\text{A}(\mathrm{Li})$}

\newcommand{\numCatalogSpect}{795,384}
\newcommand{\numCatalogStarAll}{455,752}
\newcommand{\numGALAH}{4,331}
\newcommand{\muGALAH}{0.1\,dex}
\newcommand{\sigGALAH}{0.2\,dex}
\newcommand{\numGES}{68}
\newcommand{\muGES}{0.2\,dex}
\newcommand{\sigGES}{0.3\,dex}

\shorttitle{LAMOST MRS Lithium Abundances}
\shortauthors{Ding et al.}

\graphicspath{./}

\accepted{Mar 1, 2024}

\submitjournal{ApJS}

\begin{document}

\title{Lithium Abundances from the LAMOST Med-Resolution Survey Data Release 9}

\author[0000-0001-6898-7620]{Ming-Yi Ding}
\affiliation{CAS Key Laboratory of Optical Astronomy, National Astronomical Observatories, Chinese Academy of Sciences, Beijing 100101, China}
\affiliation{School of Astronomy and Space Science, University of Chinese Academy of Sciences, Beijing 100049, China}

\author[0000-0002-0349-7839]{Jian-Rong Shi}
\affiliation{CAS Key Laboratory of Optical Astronomy, National Astronomical Observatories, Chinese Academy of Sciences, Beijing 100101, China}
\affiliation{School of Astronomy and Space Science, University of Chinese Academy of Sciences, Beijing 100049, China}
\email{sjr@nao.cas.cn}

\author[0000-0002-8609-3599]{Hong-liang Yan}
\affiliation{CAS Key Laboratory of Optical Astronomy, National Astronomical Observatories, Chinese Academy of Sciences, Beijing 100101, China}
\affiliation{School of Astronomy and Space Science, University of Chinese Academy of Sciences, Beijing 100049, China}
\affiliation{Institute for Frontiers in Astronomy and Astrophysics, Beijing Normal University, Beijing 102206, China}
\email{hlyan@nao.cas.cn}

\author[0000-0002-6647-3957]{Chun-Qian Li}
\affiliation{CAS Key Laboratory of Optical Astronomy, National Astronomical Observatories, Chinese Academy of Sciences, Beijing 100101, China}
\affiliation{School of Astronomy and Space Science, University of Chinese Academy of Sciences, Beijing 100049, China}

\author[0000-0003-4972-0677]{Qi Gao}
\affiliation{CAS Key Laboratory of Optical Astronomy, National Astronomical Observatories, Chinese Academy of Sciences, Beijing 100101, China}
\affiliation{School of Astronomy and Space Science, University of Chinese Academy of Sciences, Beijing 100049, China}

\author[0000-0002-6448-8995]{Tian-Yi Chen}
\affiliation{CAS Key Laboratory of Optical Astronomy, National Astronomical Observatories, Chinese Academy of Sciences, Beijing 100101, China}
\affiliation{School of Astronomy and Space Science, University of Chinese Academy of Sciences, Beijing 100049, China}

\author[0000-0002-2510-6931]{Jing-Hua Zhang}
\affiliation{South-Western Institute for Astronomy Research, Yunnan University, Chenggong District, Kunming 650500, China}

\author[0000-0001-5193-1727]{Shuai Liu}
\affiliation{CAS Key Laboratory of Optical Astronomy, National Astronomical Observatories, Chinese Academy of Sciences, Beijing 100101, China}

\author[0000-0002-4440-4803]{Xiao-Jin Xie}
\affiliation{CAS Key Laboratory of Optical Astronomy, National Astronomical Observatories, Chinese Academy of Sciences, Beijing 100101, China}
\affiliation{School of Astronomy and Space Science, University of Chinese Academy of Sciences, Beijing 100049, China}

\author{Yao-Jia Tang}
\affiliation{CAS Key Laboratory of Optical Astronomy, National Astronomical Observatories, Chinese Academy of Sciences, Beijing 100101, China}
\affiliation{School of Astronomy and Space Science, University of Chinese Academy of Sciences, Beijing 100049, China}

\author[0000-0002-1619-1660]{Ze-Ming Zhou}
\affiliation{Institute for Frontiers in Astronomy and Astrophysics, Beijing Normal University, Beijing 102206, China}
\affiliation{Department of Astronomy, Beijing Normal University, Beijing 100875, China}

\author[0000-0002-2316-8194]{Jiang-Tao Wang}
\affiliation{CAS Key Laboratory of Optical Astronomy, National Astronomical Observatories, Chinese Academy of Sciences, Beijing 100101, China}

\begin{abstract}
Lithium is a fragile but crucial chemical element in the universe, exhibits interesting and complex behaviors. 
Thanks to the massive spectroscopic data from the Large Sky Area Multi-Object Fiber Spectroscopic Telescope (LAMOST) medium-resolution survey (MRS), we can investigate the lithium abundances in a large and diverse sample of stars, which could bring vital help to study the origin and evolution of lithium. 
In this work, we use the \ion{Li}{1} 6,707.8\,\AA\ line to derive the lithium abundance through a template-matching method.
A catalog of precise lithium abundance is presented for \numCatalogSpect\ spectra corresponding to \numCatalogStarAll\ stars from the LAMOST MRS Data Release (DR) 9.
Comparing our results with those of external high-resolution references we find a good consistency with a typical deviation of $\sigma \text{A}(\mathrm{Li}) \sim$ \sigGALAH.
We also analyze the internal errors using stars that have multiple LAMOST MRS observations, which will reach as low as 0.1\,dex when the signal-to-noise ratio (S/N) of the spectra $> 20$.
Besides, our result indicates that a small fraction of giant stars still exhibit surprisingly high amount of lithium contents, and 967 stars are identified as Li-rich giants with A(Li) $>1.5$\,dex, accounting for $\sim 2.6\%$ of our samples. 
If one takes into account the fact that nearly all stars deplete lithium during the main sequence, then the fraction of Li-rich stars may exceed 2.6\% much. 
This new catalog covers a wide range of stellar evolutionary stages from pre-main sequence to giants, and will provide help to the further study of the chemical evolution of lithium.

\end{abstract}


\section{Introduction} \label{sec:intro}
The Standard Big Bang Nucleosynthesis (SBBN) theory suggests that lithium, the heaviest element formed just after the Big Bang \citep{wagoner1973, yang1984, olive1990, malaney1993, mathews2017, randich2021}, plays a crucial role in unveiling the early stage of the universe \citep{Coc2014StandardBigBang}.
Currently, based on the baryon-to-photon ratio suggested by interpretation of measurements of the microwave background radiation, SBBN predicts robustly a primordial lithium abundance of \ALi\ $= 2.7$\,dex \citep{Cyburt2003PrimordialNucleosynthesisLight, Coc2012STANDARDBIGBANG, Coc2014CanMirrorMatter, Coc2014StandardBigBanga, Cyburt2008UpdateBigBang, Cyburt2016BigBangNucleosynthesis, Fields2020BigBangNucleosynthesisPlanck, Spergel2003FirstYearWilkinson}. 
However, the long observation of lithium abundance in the metal-poor halo stars near turnoff contradicts this prediction. 
In particular, \citet{Spite1982AbundanceLithiumUnevolved} found that the metal-poor unevolved stars of \teff $>$ 5,500\,K exhibit a plateau of lithium abundance, which is independence with effective temperature, referred to as the Spite-Plateau in \citet{Deliyannis1990}.
Similarly, a plateau is clearly detected by \citet{Rebolo1988}, and they suggested an independence of \ALi\ with metallicity for metal-poor dwarfs. The primordial lithium abundance exhibited by the Li plateau is around \ALi\ $\sim 2.2$\,dex, which is more than three times lower than that of the SBBN prediction.

The severe lithium depletion phenomenon in the universe is called cosmological lithium problem.
For years, various solutions have been supposed \citep[e.g.,][]{Endal1976EvolutionRotatingStars, Endal1978EvolutionRotatingStars, Endal1981RotationSolartypeStars, Pinsonneault1989EvolutionaryModelsRotating, Pinsonneault1992, Chaboyer1994, descouvemont2004, kusakabe2015, hou2017, kang2019, korn2020, sasankan2020}.
For example, \citet{Deliyannis1990} described how the choice of particular parameters leads to a factor of 3 lithium depletion in halo dwarfs \citep[as opposed to the factor of 10 predicted by][]{Pinsonneault1992, Chaboyer1994}, long before it was suggested by studies of primordial D and microwave background that a factor of 3 lithium depletion would be needed to make SBBN self-consistent at the same baryon-to-photon ratio; also, \citet{Pinsonneault1999} investigated the stellar models with rotational mixing and suggested a depletion range of 0.2--0.4\,dex, which is close to the SBBN needed value.
Additionally, the present lithium abundance is found to be \ALi\ $=3.28$\,dex in meteorites \citep{Lodders2009AbundancesElementsSolar}, significantly exceeding the solar abundance and the SBBN prediction, which suggests that lithium has undergone some enrichment mechanism, despite being consumed in the interior of stars. 

The evolution of lithium in the Milky Way presents a major challenge in modern astrophysics.
On the one hand, lithium is easily destroyed by nuclear burning in stellar interiors \citep{Gamow1933InternalTemperatureStars, Salpeter1955NuclearReactionsStars}; on the other hand, observations show that lithium can be produced at some specific evolutionary stages. 

Presently, growing observational evidence indicates that the thin- and thick-disk stars may have experienced different lithium enrichment paths. \citet{Molaro1997LithiumVeryMetalpoor} noted that the metal-poor thick-disk stars exhibit the same lithium abundance as the Spite-Plateau lithium abundance found in halo stars. 
\citet{Guiglion2016AMBREProjectConstraining} constructed a homogeneous catalog of lithium abundances for 7,300 stars and found that lithium in the thick-disk increases slightly with metallicity, while thin-disk stars exhibit a steeper increase of lithium enrichment with metallicity. 
The different lithium enrichment behaviors between the thin and thick disk is likely due to two possibilities: the different Galactic lithium production between the two groups, and/or the different stellar lithium depletion between the two groups, for example, the different distributions in initial angular momentum between the two groups could lead to different lithium depletion \citep[see in details,][]{Pinsonneault1992}.

In young open clusters, lithium content shows a remarkable dependence on effective temperature (\teff) \citep{Boesgaard1986LithiumHyadesCluster, Soderblom1993EvolutionLithiumAbundances}. 
Hotter stars (\teff\ $>$ 6,800\,K), with relatively shallow convective zones, which should preserve the primordial lithium within their stellar atmospheres that have not been fully mixed down to high-temperature interiors however, they deplete lithium over time even as they spin down \citep{Deliyannis2019LiEvolutionOpen}.
A different mechanism must be at work, as the lithium depletion and spin-down is correlated \citep[see the Yale rotational models, e.g.,][]{Endal1976EvolutionRotatingStars, Endal1978EvolutionRotatingStars, Endal1981RotationSolartypeStars, Pinsonneault1989EvolutionaryModelsRotating}.
The cooler stars (\teff\ $<$ 5,800\,K) experience severe lithium depletion, and the standard theory can explain that cooler dwarfs have deeper surface convection zones, which result in a greater lithium depletion \citep{Deliyannis1990}, however, this process only takes place during the pre-main sequence (PMS). 
It is found that the open clusters older than the Pleiades show lithium depletion in G (and cooler) dwarfs, and the depletion continues during the main sequence (MS), resulting to the lithium abundances much lower than those seen in the Pleiades. 
The most likely mechanism is rotational mixing, and evidences suggest that this lithium depletion correlates with spin-down \citep{Pinsonneault1990RotationLowMassStars}, and especially as evidenced by the lithium abundance observed in Short Period Tidally Locked Binaries \citep{Ryan1995LithiumShortPeriodTidallya, Thorburn1993LithiumHyadesNew}.
In the mid-temperature range, roughly between 6,500 and 6,850\,K, where lithium content is rapidly depleted within this narrow temperature interval resulting in what is known as the Li-Dip phenomenon, where different kinds of evidence supporting rotational mixing as the dominant mechanism that creates the Li-Dip \citep[e.g.,][]{Jeffries1997MembershipLithiumAbundances, Boesgaard1986LithiumHyadesCluster, Jeffries2002MembershipMetallicityLithium, Somers2015RotationInflationLithium, Somers2015OLDERCOLDERIMPACT, Deliyannis2019LiEvolutionOpen}.
In addition, the Li-Dip is no longer as narrow as originally suspected by \citet{Boesgaard1986LithiumHyadesClustera}, in fact, it has a considerable structure. 
More specifically, there is a wall (or the hot or blue side of the Li-Dip) of lithium abundances ranging over two \,dex near 6,800--6,700\,K \citep{AnthonyTwarog2009LITHIUMINTERMEDIATEAGEOPEN, Deliyannis2019LiEvolutionOpen, Twarog2020IntermediatetolowMassStars, Boesgaard2022LithiumBerylliumNGC, Sun2023WIYNOpenCluster}, then a very deep part near 6,700--6,600\,K (the original Li-Dip), after them the lithium abundance increases for stars from 6,600\,K all the way to 6,300\,K.
Also, there is a Li plateau (6,300--5,900\,K) between the Li-Dip and the depletion for \teff\ $<$ 5,800\,K.
Note that this plateau also depletes over time, as evidenced by SPTLBs \citep{Deliyannis1994LithiumShortPeriodTidally} and comparison of clusters of the same metallicity with different ages \citep{Boesgaard2022LithiumBerylliumNGC}, which might provide possible connection to the Spite-Plateau.

According to the standard stellar evolutionary model, the surface lithium depletion could occur during the PMS and early MS via the nuclear burning at the base of the convection zone \citep{Proffitt1989PreMainSequenceDepletion, Deliyannis1990LithiumHaloStarsa, Deliyannis1991LithiumMostExtreme}.
Nearly all dwarfs in which the lithium line can be measured, which includes all of A, F, G, K, and M dwarfs, deplete their lithium over time.
In particular, the lithium depletion for MS stars in open clusters such as Hyades and Praesepe, are in clear contradiction to standard models \citep{Cummings2017WIYNOpenCluster}.
The spin-down of MS stars could lead to angular momentum loss, which relates to rotational mixing, such as the Yale-style and other models based on these or related precepts are considered as relevant factors of this phenomenon \citep{Pinsonneault1990RotationLowMassStars, Steinhauer2004WIYNHydraDetection, Deliyannis2019LiEvolutionOpen}.
Also, the observations of Beryllium (Be) and Boron (B), they can survive to deeper layers, could be an important evidence that rotational mixing depletes lithium in FGK dwarfs \citep{Deliyannis1998CorrelatedDepletionLithiuma, Boesgaard2004CorrelationLithiumBeryllium, Boesgaard2005BoronDepletionDwarf, Boesgaard2020CorrelatedDepletionDilution}.
\citet{Somers2015RotationInflationLithium} investigated the Li-rotation correlation in the Pleiades, and showed an evidence that magnetic fields in rapid rotators can help inflate stellar radii and preserve lithium \citep{Jackson2018InflatedRadiiDwarfs, Jackson2019SearchRadiusInflation}.

The first dredge-up (FDU) process, occurring as stars evolve off the MS, can further bring the surface lithium into deeper layers of the convective zone.
At this stage, lithium can be severely destroyed via nuclear burning, as a result, a widely suggested upper limit is predicted as \ALi\ $<$ 1.5\,dex for red giant branch (RGB) stars \citep[e.g.,][]{Iben1965StellarEvolutionII, Iben1967StellarEvolutionVI, Brown1989SearchLithiumrichGianta, Kumar2011OriginLithiumEnrichment, Kumar2018IdentifyingLirichGiants}. 
Besides, there are several non-standard mechanisms suggesting lithium depletion during a star's lifetime (e.g., stellar rotation \citep{Chaboyer1995StellarModelsMicroscopic, Charbonnel2005InfluenceGravityWavesa}, magnetic activity \citep{Denissenkov2010MODELMAGNETICBRAKING}, atomic diffusion \citep{Michaud1986LithiumAbundanceGap} and planet engulfment/accretion \citep{Galarza2021EvidenceRockyPlanet, Siess1999AccretionBrownDwarfs, DelgadoMena2015LiAbundancesStars}). 
Observations found that a small fraction of evolved stars can still retain anomalously high lithium abundance \citep{Wallerstein1982GiantUnusuallyHigh}. 
These stars, known as Li-rich giants, have lithium abundance that significantly exceeds the model prediction, indicating that there are still some unknown processes that noticeably enrich the surface lithium.
While the defination of a Li-rich giant with \ALi\ $>$ 1.5\,dex assumes that the only lithium depletion mechanism is the subgiant dilution by about 1.8 dex \citep[from meteoritic 3.3\,dex to 1.5\,dex,][]{Iben1965StellarEvolutionII, Iben1967StellarEvolutionVI, Charbonnel2010}.
However, when taking the MS lithium depletion into account, the dilution by 1.8\,dex will result in abundances lower than 1.5\,dex, or even possibly much lower \citep{Sills2000, Sun2022, Chaname2022}. 
Therefore, assuming only with \ALi\ $>$ 1.5\,dex as Li-rich giants, could miss a whole bunch of stars with lower \ALi\ that have more lithium contents than expected from the combination of MS lithium depletion and subgiant dilution, therefore, the fraction of Li-rich stars will be underestimated.

To better understand the complex behavior of lithium in the universe, a catalog of lithium abundances derived from a homologous and consistent spectroscopic survey is necessary. 
Over the decades, considerable efforts have been dedicated to obtain precise lithium abundances, essential for unraveling the mechanisms of lithium enrichment and depletion. 
Spectroscopic surveys, such as the Large Sky Area Multi-Object Fiber Spectroscopic Telescope \citep[LAMOST,][]{Cui2012LargeSkyArea, Yan2022OverviewLAMOSTSurvey}, the \textit{Gaia}-ESO \citep{Gilmore2012GaiaESOPublicSpectroscopic}, the GALactic Archaeology with HERMES \citep[GALAH,][]{Silva2015GALAHSurveyScientific, Buder2018GALAHSurveySecond, Buder2021GALAHSurveyThirda} and the AMBRE Project \citep{DeLaverny2013AMBREProjectStellar}, have provided vast number of valuable low- and high-resolution stellar spectra, enabling us to deduce lithium contents for a large and diverse sample of stars. 
For example, \citet{Franciosini2022GaiaESOSurveyLithium} measured lithium equivalent widths (EWs) and abundances of $\sim$ 40,000 stars from the \textit{Gaia}-ESO survey, predominantly distributed in open clusters.
\citet{martos2023} employed high-resolution HARPS spectra and analyzed the impact of metallicities, ages and planets on lithium abundances for 41 solar analogues.
\citet{Gao2021LithiumAbundancesLarge} derived a large sample of lithium abundances for approximately 300,000 LAMOST spectra. 
Base on the LAMOST low-resolution survey, \citet{zhou2018} investigated a super Li-rich (\ALi\ $>=$ 3.3\,dex, the typical interstellar medium lithium abundance, much more rare than Li-rich giants \citep[e.g.,][]{DeLaReza1995, Balachandran2000, Kumar2009, Adamow2015}) K giant with a low carbon isotopic ratio and suggested that extra mixing can induce Cameron-Fowler mechanism which enhanced lithium in this star.
Similarly, \citet{yan2022b} discovered 9 unevolved stars with unusually high levels of lithium abundances (\ALi\ $>$ 3.8\,dex) from the LAMOST med-resolution survey, which indicates a different history of lithium enrichment for those unevolved stars.
\citet{kowkabany2022} reported the discovery of a very metal-poor (VMP), ultra Li-rich giant with $\text{A}(\mathrm{Li})_{\text{(3D, NLTE)}} = 5.62$\,dex and investigated the possibility of lithium production.
Additionally, by combining the LAMOST spectra with the asteroseismic measurements from the \textit{Kepler} mission \citep{Borucki2010KeplerPlanetDetectionMission},
 \citet{yan2020a} presented a study of lithium abundances between the red giant branch (RGB) and the red clump (RC) stars, which revealed that the RC stars are more frequently Li-rich rather than traditionally supposed RGB stars. 
\citet{zhang2021d} further investigated the evolutionary features of lithium for stars evolved from RGB to He-burning (HeB) phase and suggested that the helium flash can be responsible for moderate lithium production.
Likewise, \citet{bandyopadhyay2022} presented a study on the lithium distribution and the kinematics of VMP stars with astrometric parameters from the \textit{Gaia} mission \citep{gaiacollaboration2021} and the high-resolution spectra observed from the Hanle Echelle SPectrograph - Galactic survey Of metal-poor stArs \citep[HESP-GOMPA,][]{kumar2018Hanle} survey. 

In this paper, we present our work in the following structure. Section~\ref{sec:obs} describes the spectra observed by the LAMOST survey and our sampling strategy. Section~\ref{sec:met} introduces the theoretical spectra and the template-matching methodology. In Section~\ref{sec:valid}, we compare our results with the lithium measurements from external surveys. We discuss the deviation and estimate the typical errors in Section~\ref{sec:acc}. In Section~\ref{sec:cat}, the catalog and properties of our lithium abundances are presented. Finally, we briefly summarize our results in Section~\ref{sec:con}.

\section{Observational Data} \label{sec:obs}
The dataset used in this work are composed of two parts: the observational spectra from the LAMOST MRS DR9 and the reference set consists of lithium abundances provided by other surveys (GALAH and \textit{Gaia}-ESO) which is used for validation of our result.

\subsection{LAMOST MRS Observation} \label{subsec:obs-mrs}

The LAMOST survey uses a large-aperture reflective Schmidt telescope located at the Xinglong Observatory in Hebei province, China. 
Its focal plane contains 4,000 fibers, connected to 16 spectrographs equipped with 32 CCD cameras (4K $\times$ 4K).
In the year 2017, the spectrographs were upgraded to support two different modes: low-resolution (R$\sim$1,800) and medium-resolution (R$\sim$7,500). 
For the medium-resolution mode, the wavelength range is from 4,950 to 5,350\,\AA\ in the blue band, while the red band covers 6,300 to 6,800\,\AA.

The ninth data release (DR9) of the LAMOST survey includes 8,259,362 medium-resolution and 11,211,028 low-resolution spectra, respectively. 
The large coverage of the stellar samples observed by LAMOST makes it possible to obtain a homogeneous large sample set of lithium abundances.

\subsection{Stellar Parameters} \label{subsec:para}
In addition to the vast number of spectra mentioned above, LAMOST DR9 also equips us with stellar parameter measurements through the LAMOST Stellar Parameter Pipeline \citep[LASP,][]{Luo2015FirstDataRelease}. LASP utilizes an empirical spectral library \citep[ELODIE,][]{Prugniel2001DatabaseHighMediumresolution} and employs a template-matching method to derive the stellar parameters.  
Furthermore, we cross-match our data with the Apache Point Observatory Galactic Evolution Experiment \citep[APOGEE,][]{Majewski2017ApachePointObservatory} DR17.
APOGEE is one of the subprojects of the Sloan Digital Sky Survey (SDSS) \citep{Eisenstein2011SDSSIIIMassiveSpectroscopic}, which offers high-resolution (R $\sim$ 22,500), high signal-to-noise ratio (S/N $>$ 100), near-infrared spectra for approximately 650,000 stars. 
The APOGEE Stellar Parameter and Chemical Abundances Pipeline \citep[ASPCAP][]{GarciaPerez2016ASPCAPAPOGEESTELLAR} fits these observational spectra through comparison to the precomputed grids \citep{Gustafsson2008GridMARCSModel} and provides stellar parameter measurements with a local thermodynamic equilibrium (LTE) assumption \citep{Jonsson2020APOGEEDataSpectral}.

\subsection{Data Preparation}
\label{subsec:datapre}
In order to obtain the synthetic spectra, we generate a series of similar templates of the same stellar parameters (\teff, \logg, \feh\ and micro-turbulence velocity $\xi$) with different lithium abundances, so that the $\chi^2$ minimization can be performed on the targets. 
To achieve this goal, we first use the stellar parameters provided by LAMOST LRS DR9 for each MRS spectrum. 
Then adopt the APOGEE measurements for the common stars to increase the number of our samples. 
For the stars that have both LASP and ASPCAP stellar parameters, we keep the ASPCAP value since the latter is obtained from high-resolution spectra.

\begin{figure*}[!htbp]
	\centering
	\gridline{
		\fig{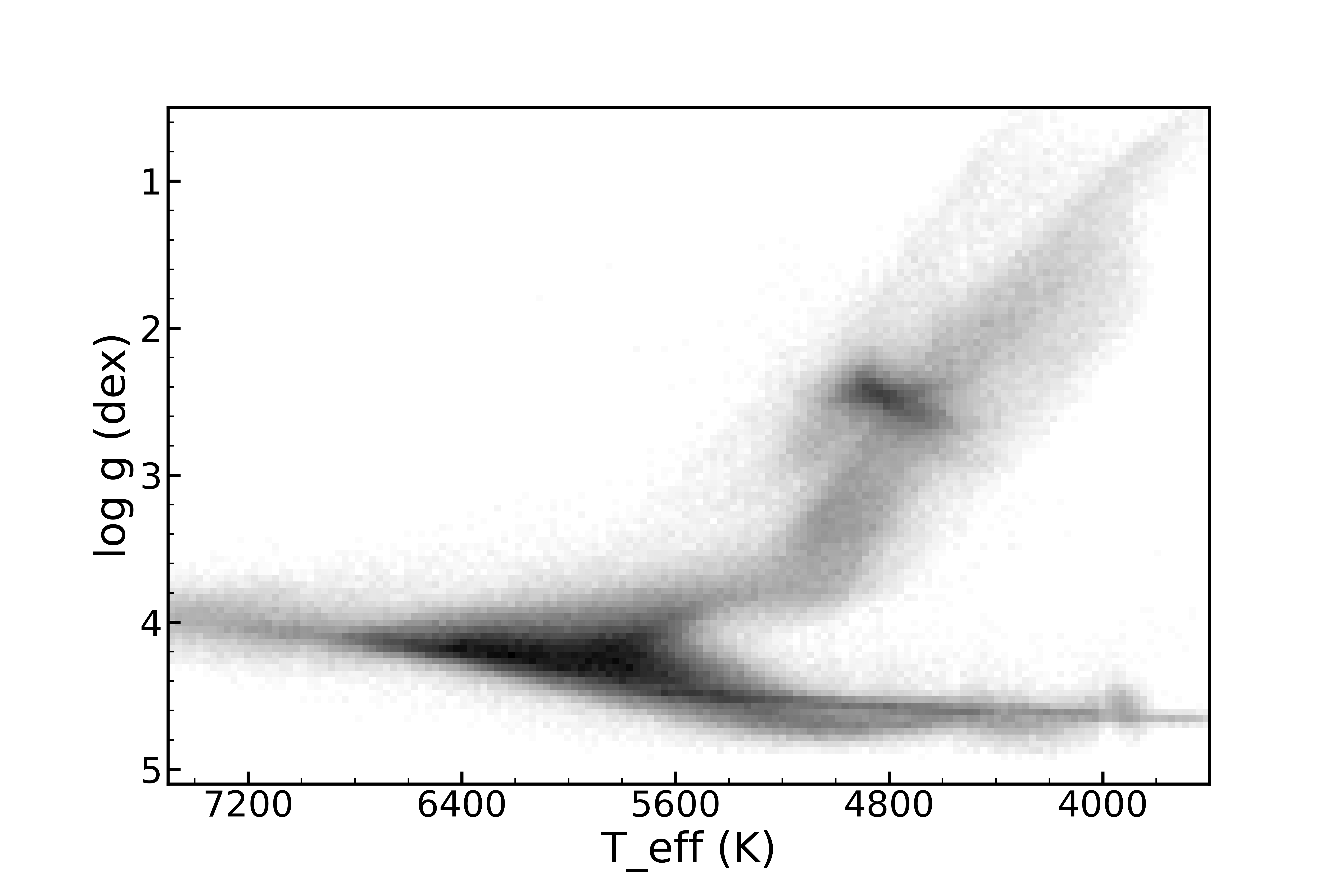}{0.35\textwidth}{}
		\fig{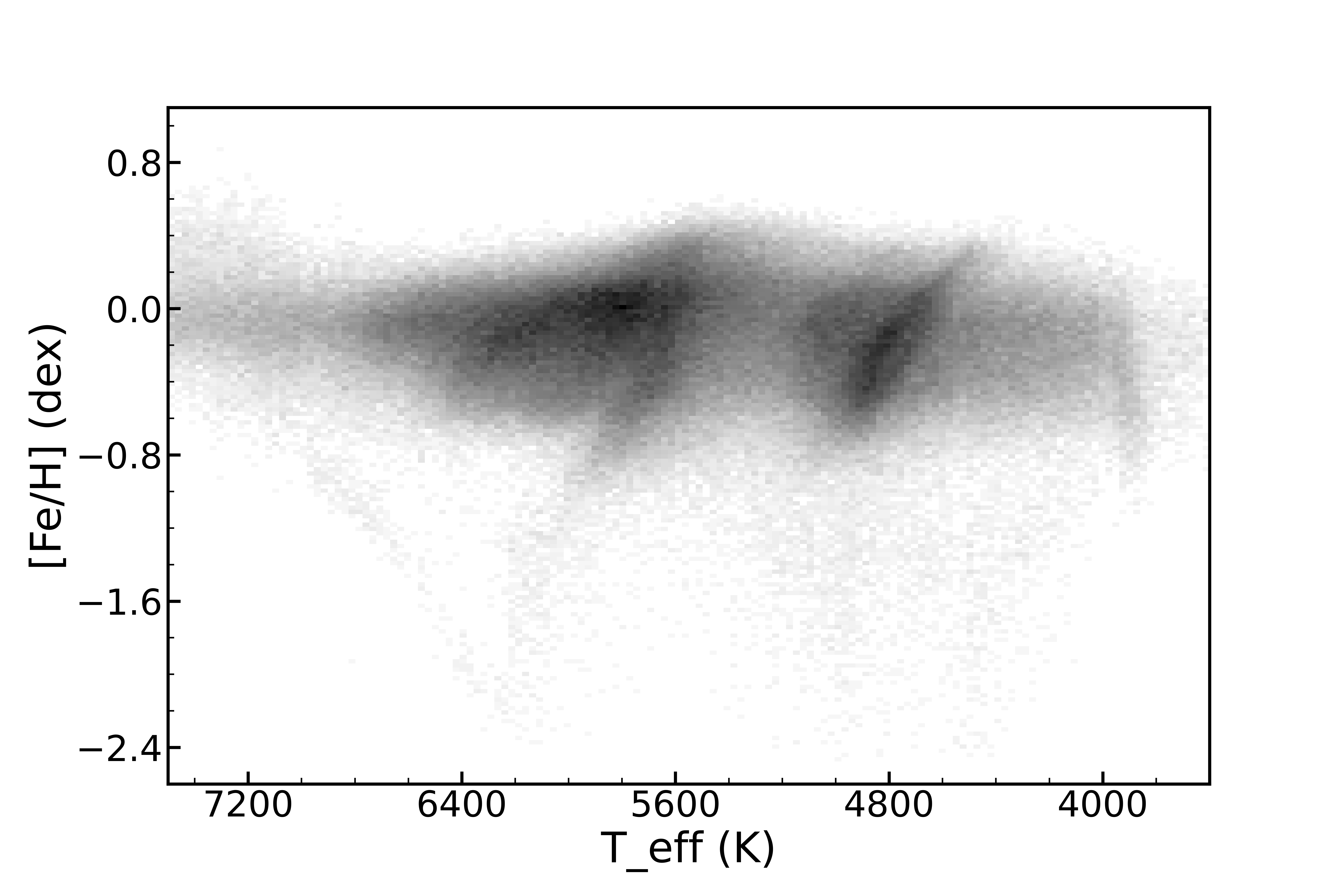}{0.35\textwidth}{}
		\fig{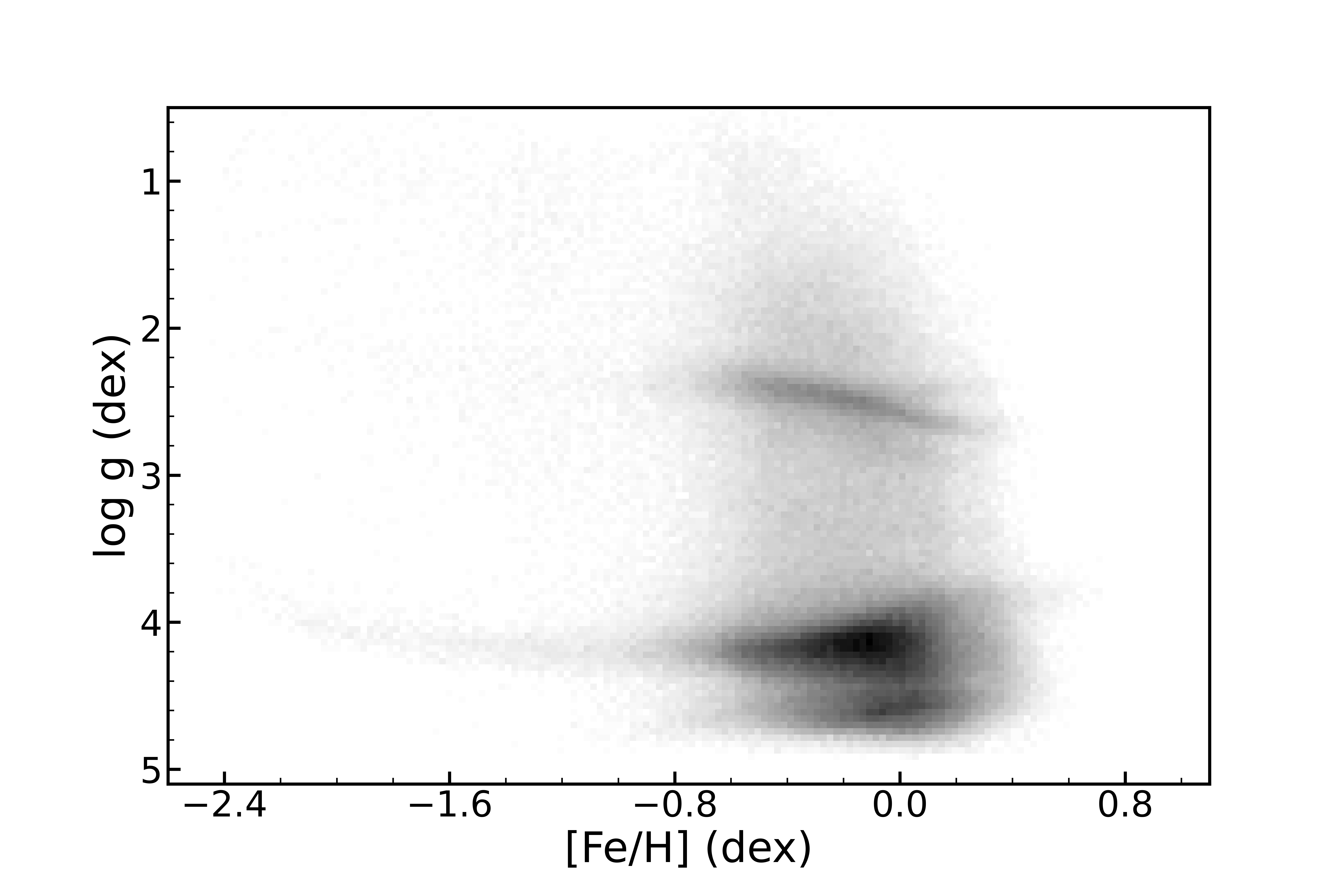}{0.35\textwidth}{}}
	\caption{The distributions of stellar parameters for our samples.}
	\label{fig:HR_diagram_1}
\end{figure*}

Finally, we acquire a sample set of \numCatalogStarAll\ individual stars with both stellar parameters and LAMOST MRS spectra.
An illustration of the parameter space distribution of our samples is shown in Figure.~\ref{fig:HR_diagram_1}.

\section{Methodology} \label{sec:met}
The template matching method is commonly adopted for determining the stellar parameters in spectral research, which searches for the minimum $\chi^2$ between the observed spectra and the templates. 

\citet{Xiang2015LAMOSTStellarParameter} used a template matching method to exploit the LAMOST Stellar Parameter Pipeline at Peking University (LSP3) to determine radial velocities (RV) and stellar atmospheric parameters for the LAMOST survey. 
Utilizing the template-matching technique of the LSP3 pipeline, \citet{Li2016MethodMeasuringFe} has derived the [$\alpha$/Fe] ratios from the LAMOST low-resolution spectra. 
\citet{Gao2019LithiumrichGiantsLAMOST} adopted a similar approach to measure [Li/Fe] and search for Li-rich giants from the LAMOST DR7. 
In this study, we revised the method similar to \citet{Gao2021LithiumAbundancesLarge} for lithium abundance measurement. 

\subsection{The templates} \label{subsec:templates}
In order to derive the lithium abundance from the LAMOST MRS spectra, we use a grid of synthetic spectra which were computed with the SPECTRUM synthesis code\footnote{\url{https://www.appstate.edu/~grayro/spectrum/spectrum.html}}.
The synthetic spectra are based on the Kurucz stellar atmosphere model \citep[ATLAS9,][]{Castelli2003NewGridsATLAS9} in LTE condition with the atomic line data for lithium adopted from \citet{Shi2007LithiumAbundancesMetalpoor}.

Based on the original model grids, a finer grid of synthetic spectra that covers cold to hot stars is calculated with the stellar atmospheric parameters as follows:

\begin{itemize}
	\item[] \teff\ (cold stars) : 3,500--4,500\,K, step 200\,K;
	\item[] \teff\ (warm stars) : 4,500--6,500\,K, step 100\,K;
	\item[] \teff\ (hot stars) : 6,500--7,500\,K, step 200\,K;
	\item[] \logg : 0.0--5.0\,dex, step 0.1\,dex;
	\item[] \feh : $-2.6$--0.6\,dex, step 0.2\,dex;
	\item[] [Li/Fe] : $-3.0$--5.1\,dex, step 0.1\,dex.
\end{itemize}

The micro-turbulence velocity is important in the lithium measurement, in this work, we calculate the micro-turbulent velocity separately for different stellar samples using a series of empirical relations \citep[see][for detail information]{Gao2021LithiumAbundancesLarge}.
The wavelength range of the synthetic spectra is generated from 6,675\,\AA\ to 6,740\,\AA\ in steps of 0.1\,\AA, and the resolution will be adjusted to match the LAMOST med-resolution spectra. 

\subsection{The lithium abundances} \label{subsec:lithiumabundances}

\begin{figure*}[!htbp]
	\centering
	\gridline{
		\fig{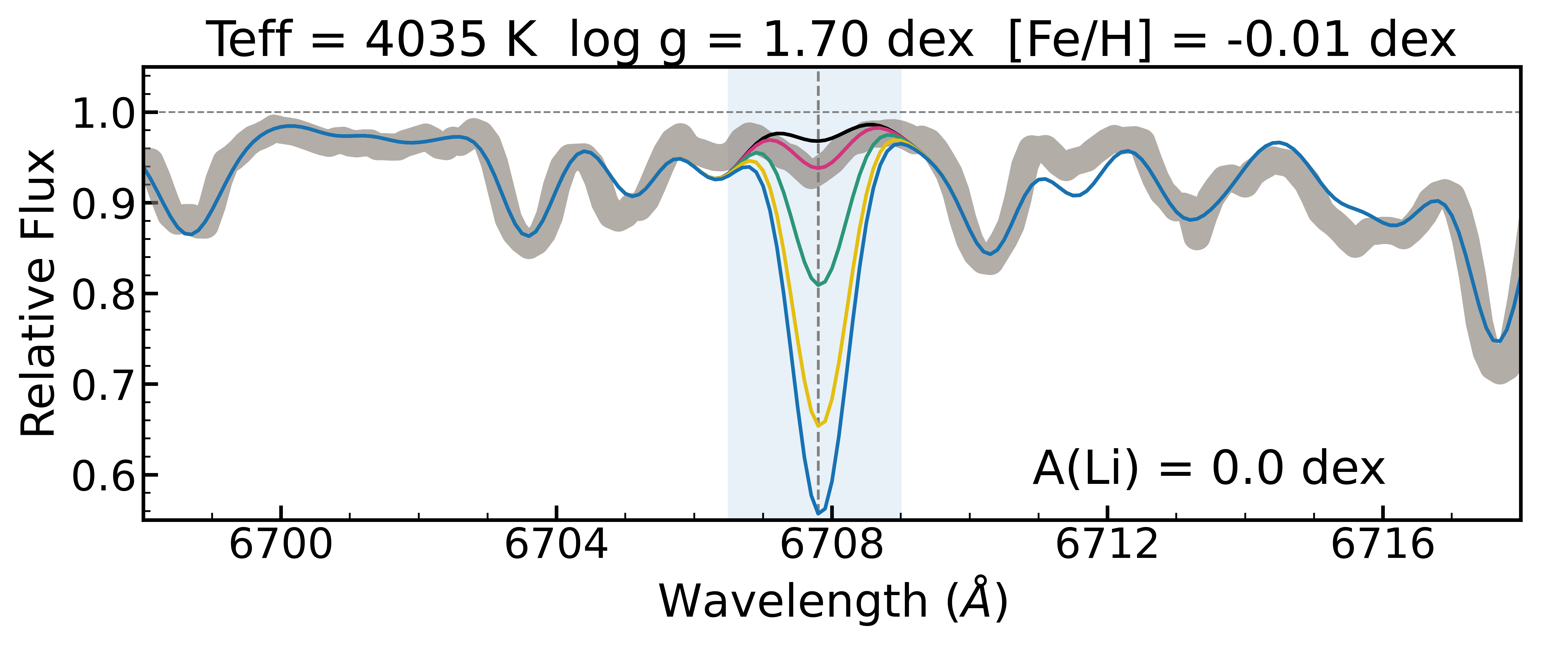}{0.5\textwidth}{}
		\fig{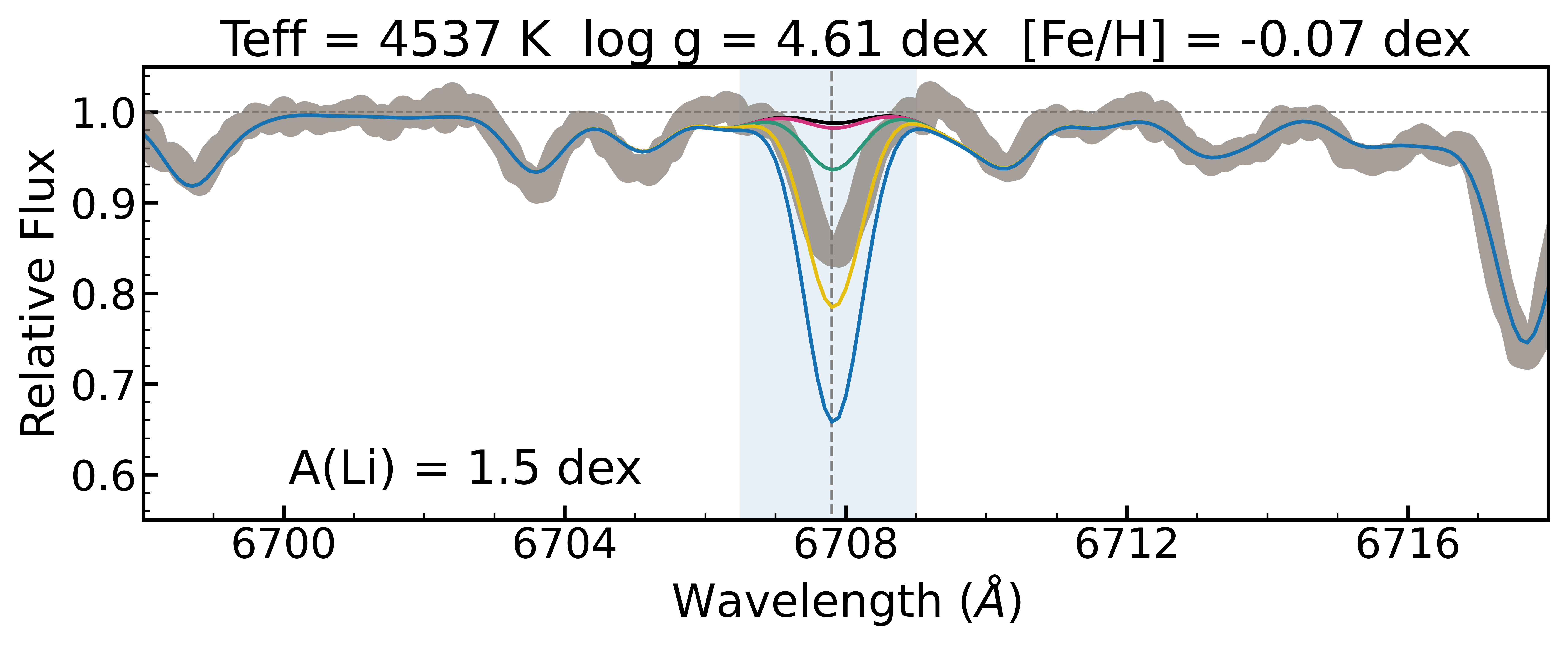}{0.5\textwidth}{}} \vspace{-7.5mm}
        \gridline{
		\fig{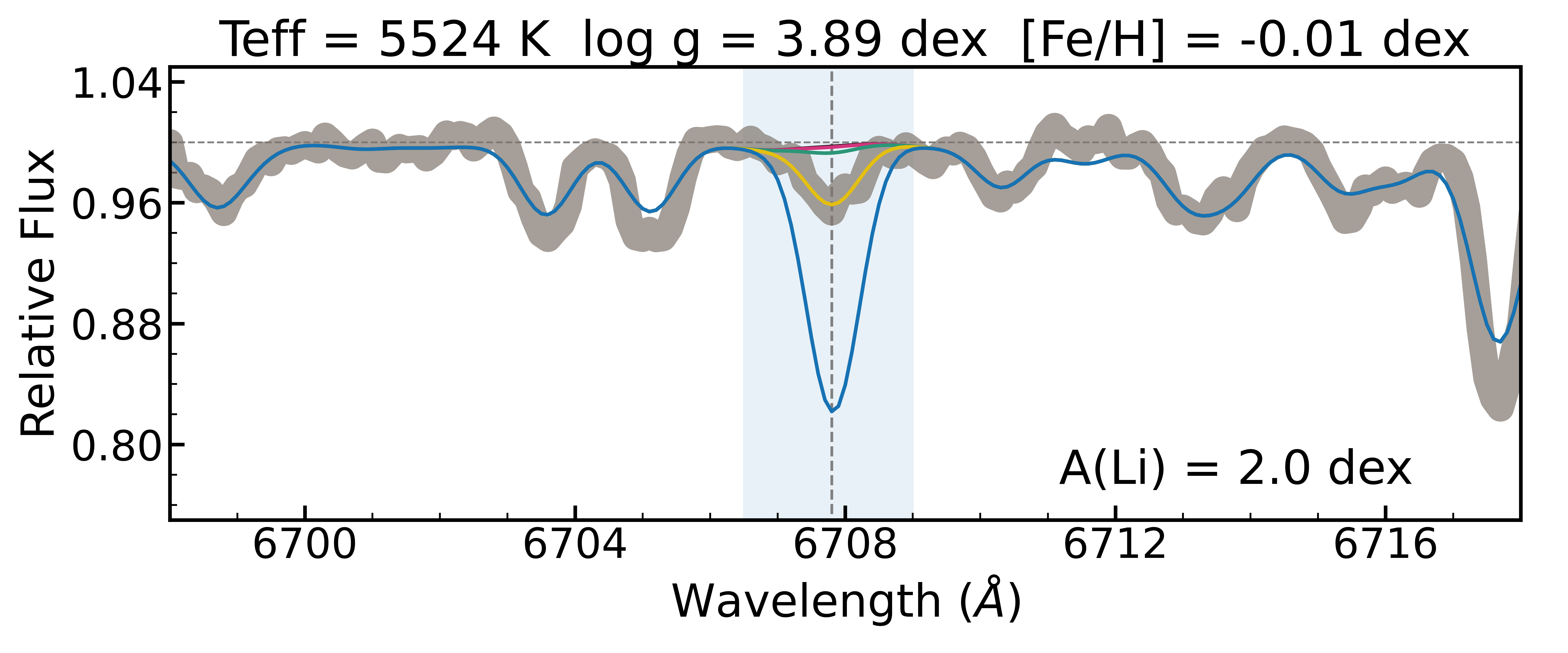}{0.5\textwidth}{}
		\fig{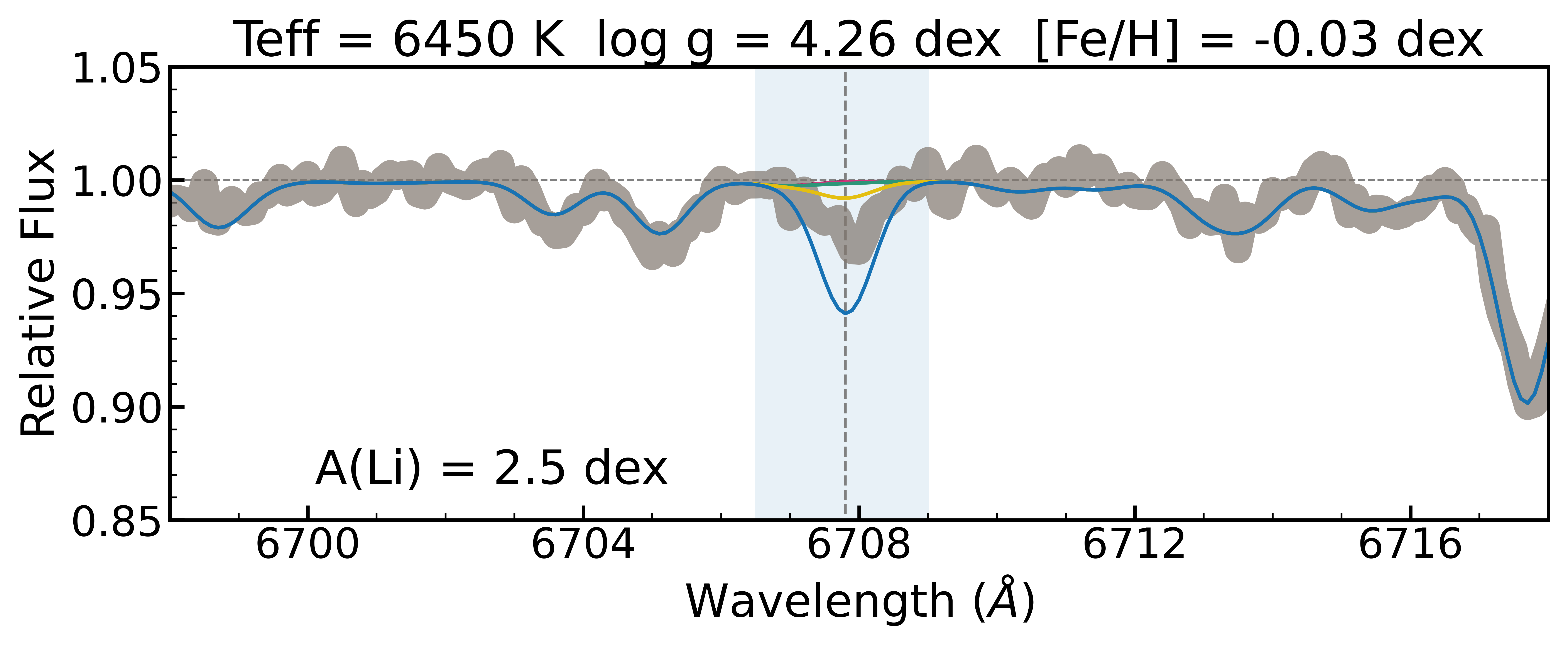}{0.5\textwidth}{}}  \vspace{-5mm}
	\caption{Examples of our fitting results for cool giant (top left), cool dwarf (top right), warm giant (bottom left), and hot dwarf (bottom right). 
	In each plane, the observed spectrum is plotted in shaded gray region, in the same time, the template spectra are plotted with \ALi$=-1.0$ (black line), 0.0 (red line), 1.0 (green line), 2.0 (yellow line), and 3.0 (blue line), respectively.
	The stellar parameters and lithium abundances are listed, and the vertical gray dotted lines show the position of \ion{Li}{1} resonance line.}
	\label{fig:MRSspec}
\end{figure*}

Before matching the spectra, we correct the wavelength of the synthetic spectra utilizing the method similar to \citet{Li2021DoubleTriplelineSpectroscopic}, which calculates the radial velocity based on the Cross Correlation Function (CCF) and successive derivatives. 

To measure the lithium abundance, we use the \ion{Li}{1} resonance line at 6,707.8\,\AA, which is covered by the red band spectrographs (6,300--6,800\,\AA) of the LAMOST MRS. Therefore, only the red band spectra have been preserved for follow-up lithium abundance measurement.
Firstly, a grid of synthetic spectra is generated with the same stellar parameters (\teff, \logg\ and \feh) adopted in Section.~\ref{subsec:datapre} with different [Li/Fe] values.
More specifically, we adopt an interpolation algorithm to generate synthetic spectra from the templates whose stellar parameters locate on the adjacent grid points. 
For each observed spectrum with specific stellar parameters, we interpolate a series of synthetic spectra with [Li/Fe] varying from $-3.0$ to 5.1\,dex in a step of 0.1\,dex.

After obtaining the synthetic spectra with known stellar parameters and [Li/Fe], we use the $\chi^2$-minimization algorithm to find out the best-marched spectra. 
The $\chi^2$ is defined as:

$$
\chi^2 = \sum_{i=1}^{m} \frac{(F_i - E_i)^2}{E_i}
$$

Here, $F_i$ is the flux at the $i_{\text{th}}$ wavelength point of the synthetic spectrum, while the $E_i$ for the observed one. 
Theoretically, the $E_i$ follows a Poisson distribution which can also be regarded as the variance $\sigma_i^2$. 
Generally, the $\chi^2$ describes the deviation between the synthetic spectrum and the observed one. 
In other words, when the minimum of $\chi^2$ is reached, we obtain the best fit to the observed spectrum.

However, the [Li/Fe] of one observed spectrum may fall between two adjacent grid points.
Thus, we used a third-order polynomial to fit the $\chi^2$ values, and the final [Li/Fe] is determined by the spectrum corresponding to the lowest $\chi^2$.  
Finally, we use the \ALi\ to indicate the lithium abundance by means of the formula : $\text{A}(\mathrm{Li}) = \mathrm{[Li/Fe]} + \mathrm{[Fe/H]} + \text{A}(\mathrm{Li})_{\odot}$.

\subsection{The Detection Efficiency} \label{subsec:upperlimit}

Figure.~\ref{fig:MRSspec} shows the four examples of our template-matching result. 
In each panel, the observed spectrum is marked in gray curve, while the templates with five different lithium abundances are marked by different colors. 
The figure clearly shows the decrease of depression of \ion{Li}{1} 6,707.8\,\AA\ line towards higher temperatures, which makes the lithium abundance hard to be detected for hot stars with weak \ion{Li}{1} lines.

To give out an estimation of upper limit in this scenario, we measure the depth of \ion{Li}{1} line, the average noise in the wavelength range of 6,600--6,800\,\AA\ and the dispersion level of residuals between the observed spectrum and the synthetic spectrum.   
If the depth of \ion{Li}{1} line is lower than those of the latter two, it can be easily drowned out, therefore, only the upper limit can be detected.
We visually check the synthetic spectra and give out several empirical upper limits of lithium detection for different types of stars.
Specifically, for the majority of FGK stars with \teff\ above 4,500\,K, 5,500\,K and 6,500\,K, the upper limits are around 0.0\,dex, 1.0\,dex and 2.0\,dex, respectively.

\section{Validation} \label{sec:valid}

\begin{figure*}[!htbp]
	\centering
	\gridline{
		\fig{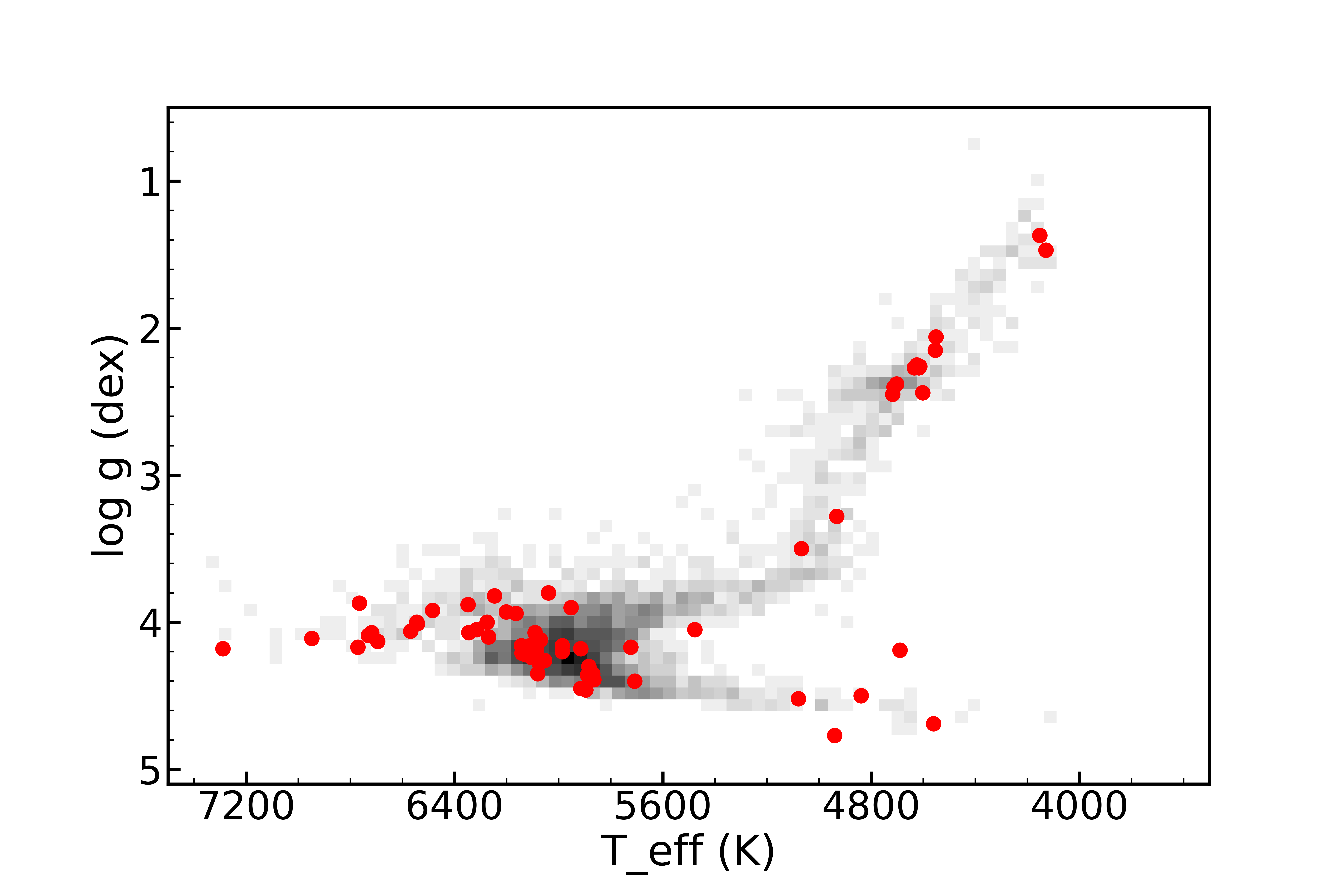}{0.35\textwidth}{}
		\fig{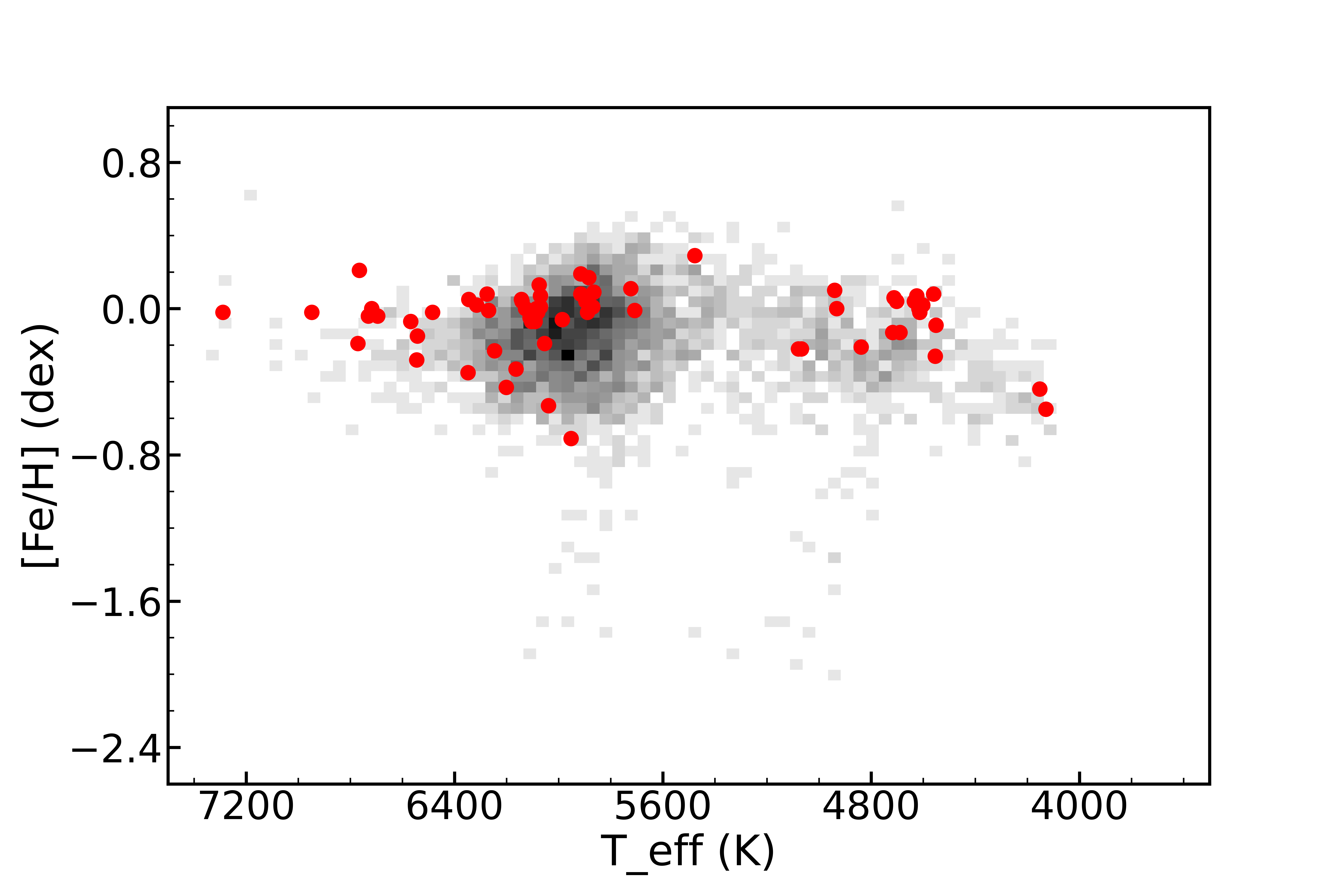}{0.35\textwidth}{}
		\fig{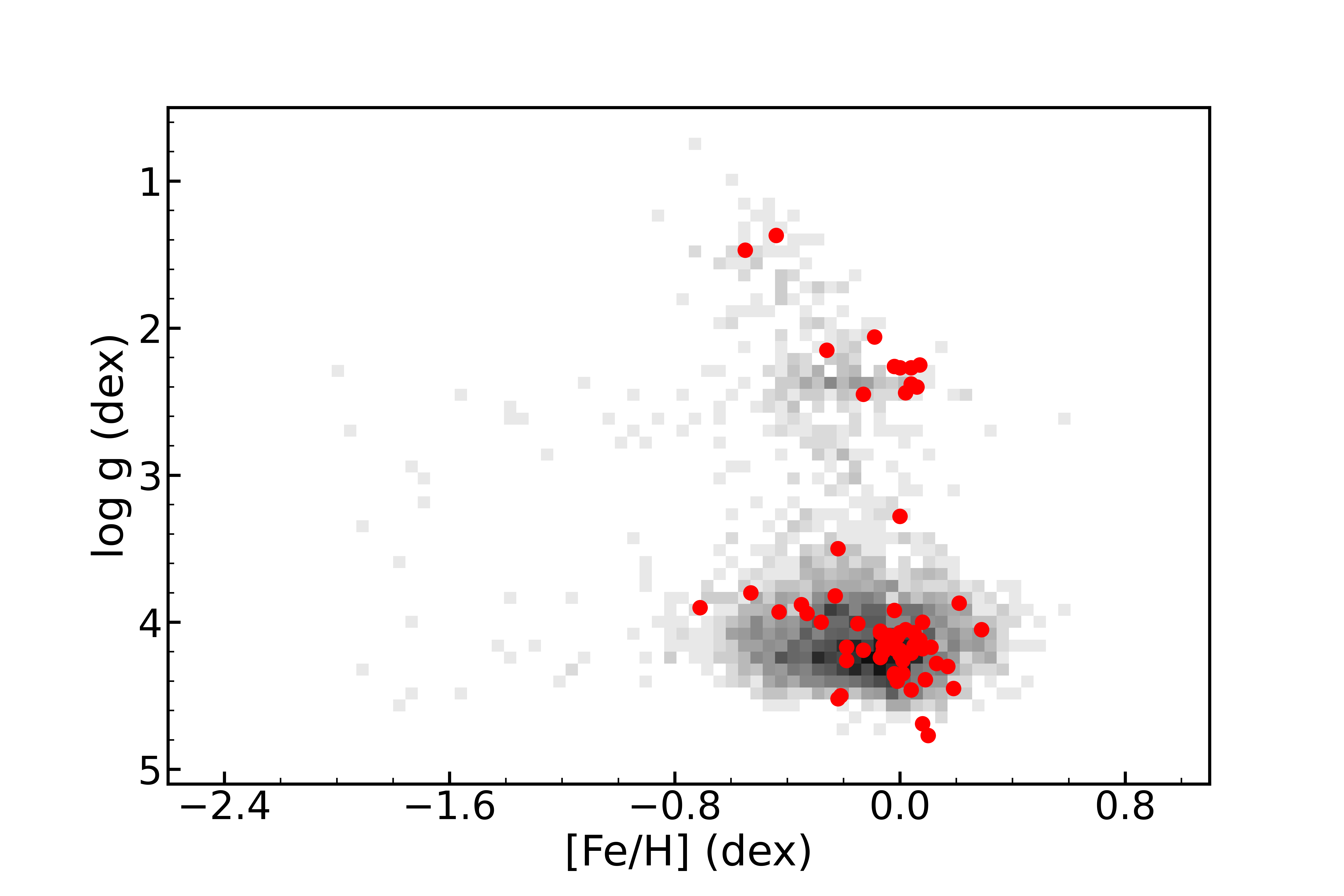}{0.35\textwidth}{}}
	\caption{The distribution of stellar parameters for common stars from the GALAH DR3 (gray distribution) and the \textit{Gaia}-ESO DR5 (red dots).}
	\label{fig:HR_diagram_GES_GALAH}
\end{figure*}

\begin{figure}[!htbp]
	\centering
		\fig{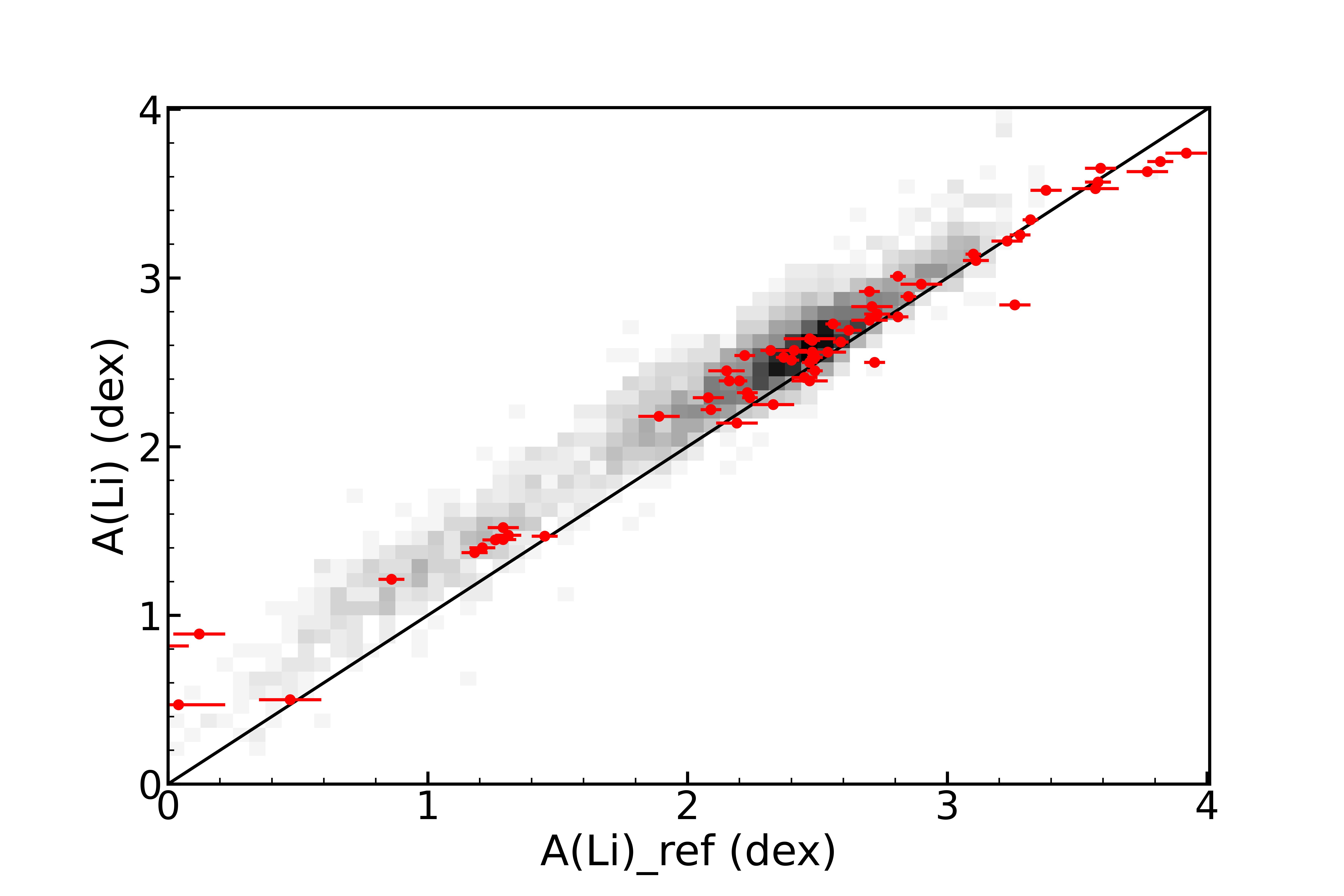}{0.5\textwidth}{}
	\caption{The comparison of the lithium abundances derived from our method and those from the GALAH DR3 and the \textit{Gaia}-ESO DR5. The error bars show the uncertainty. Colors are the same as the Figure.~\ref{fig:HR_diagram_GES_GALAH}.}
	\label{fig:LAMOSTvsRef} 
\end{figure}

\begin{figure}[!htbp]
	\centering
        \fig{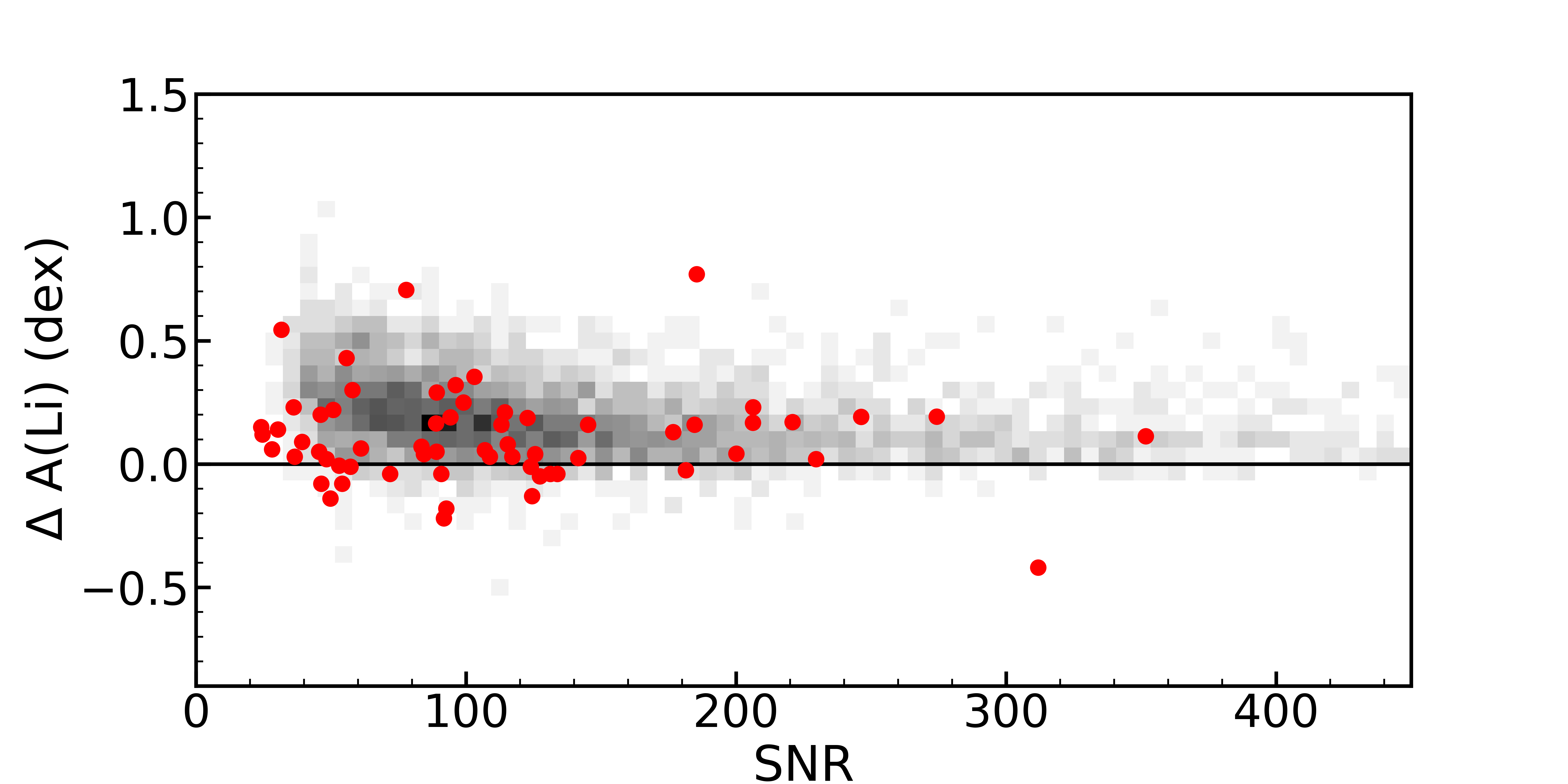}{0.5\textwidth}{}
	\caption{The residuals of the lithium abundance as a function of S/N. Colors are the same as the Figure.~\ref{fig:HR_diagram_GES_GALAH}.}
	\label{fig:ResvsRef}
\end{figure}

In order to validate the reliability and accuracy of our method, we compare our lithium abundances with those of the external measurements. 
We cross-match our samples with the GALAH DR3 \citep{Buder2021GALAHSurveyThirda} and the \textit{Gaia}-ESO DR5 \citep{Randich2022GaiaESOPublicSpectroscopic}, following the below criteria:

\begin{enumerate}
	\item{} The S/N of the LAMOST spectra should larger than 20;
	\item{} The quality flag: $\text{flag}_{\text{sp}}=0$, $\text{flag}_{\text{[Fe/H]}}=0$, $\text{flag}_{\text{[Li/Fe]}}=0$, $\text{flag}_{\text{[Li/Fe]}}=0$, and the S/N flag: $\text{snr\_c3\_iraf} > 30$ in the GALAH DR3 catalog\footnote{\url{https://www.galah-survey.org/dr3/overview/}};
	\item{} The \ion{Li}{1} measurement type: $\text{lim}_{\text{[Li1]}}=0$ in the \textit{Gaia}-ESO DR5 catalog;
	\item{} The uncertainty of [Li/Fe] as well as [Fe/H] should be lower than 0.2\,dex.
\end{enumerate}

We first exclude the spectra with S/N lower than 20 in order to affirm the reliability.
The $\text{flag}_{\text{[Li/Fe]}}$ in the GALAH DR3 catalog shows the quality of [Li/Fe] measurement. 
The objects with $\text{flag}_{\text{[Li/Fe]}}=0$ generally have more reliable lithium abundance. 
Similarly, the $\text{lim}_{\text{[Li1]}}$ in the \textit{Gaia}-ESO DR5 shows the \ion{Li}{1} measurement type, and we discard the objects with $\text{lim}_{\text{[Li1]}}=1$, which give out the upper limits of lithium abundances. 
Additionally, we use the stellar parameters (\teff, \logg\ and \feh) provided by GALAH and \textit{Gaia}-ESO instead of ours, to get rid of any errors introduced from different stellar parameters. 
In this way, we finally selected \numGALAH\ and \numGES\ common stars in the GALAH DR3 and the \textit{Gaia}-ESO DR5, respectively.
The distribution of stellar parameters for these stars is presented in Figure.~\ref{fig:HR_diagram_GES_GALAH}.

The comparison of our results with those from the common stars observed in the GALAH DR3 and the \textit{Gaia}-ESO DR5 is shown in Figure.~\ref{fig:LAMOSTvsRef}. 
Overall, good agreements between our results and those from the reference surveys can be found.
However, the comparison with GALAH indicates a systematic overestimation of $\sim$ \muGALAH\ in the lithium abundance derived from our method, while the comparison with the \textit{Gaia}-ESO demonstrates an overestimation of about \muGES\ in the same metric. 
The dispersion of \ALi\ measurements is \sigGALAH\ for the common stars in GALAH, whereas a larger spread of \sigGES\ is found for the \textit{Gaia}-ESO samples. 
Additionally, the comparison with the \textit{Gaia}-ESO shows that stars with lower \ALi\ are systematically overestimated, while those with higher \ALi\ are slightly underestimated.
However, the comparison with GALAH samples shows a rather consistent overestimation for each star.

In order to better understand how spectral quality affects the accuracy of our template-matching measurement, we explore the residuals of \ALi\ versus different S/N in Figure.~\ref{fig:ResvsRef}. 
The difference of the lithium abundance is defined as: $\Delta \text{A}(\mathrm{Li}) = \text{A}(\mathrm{Li})_{\text{LAMOST}} - \text{A}(\mathrm{Li})_{\text{Reference}}$. 
Figure.~\ref{fig:ResvsRef} shows that the $\Delta \text{A}(\mathrm{Li})$ level is consistent in all spectral quality coverage, which suggests that the overestimation actually comes from systematic bias between different methods rather than the difference in spectral qualities.

\section{Error Estimation} \label{sec:acc}
In order to evaluate the influences of the observational quality and other stellar parameters on our \ALi\ measurements, we conducted a further investigation in this section.
Thanks to the multi-observations of the LAMOST survey, we can calculate \ALi\ separately for the stars with repeated observations, and estimate the error resulting from the S/N of the spectrum.

\begin{figure}[!htbp]
	\centering
	\fig{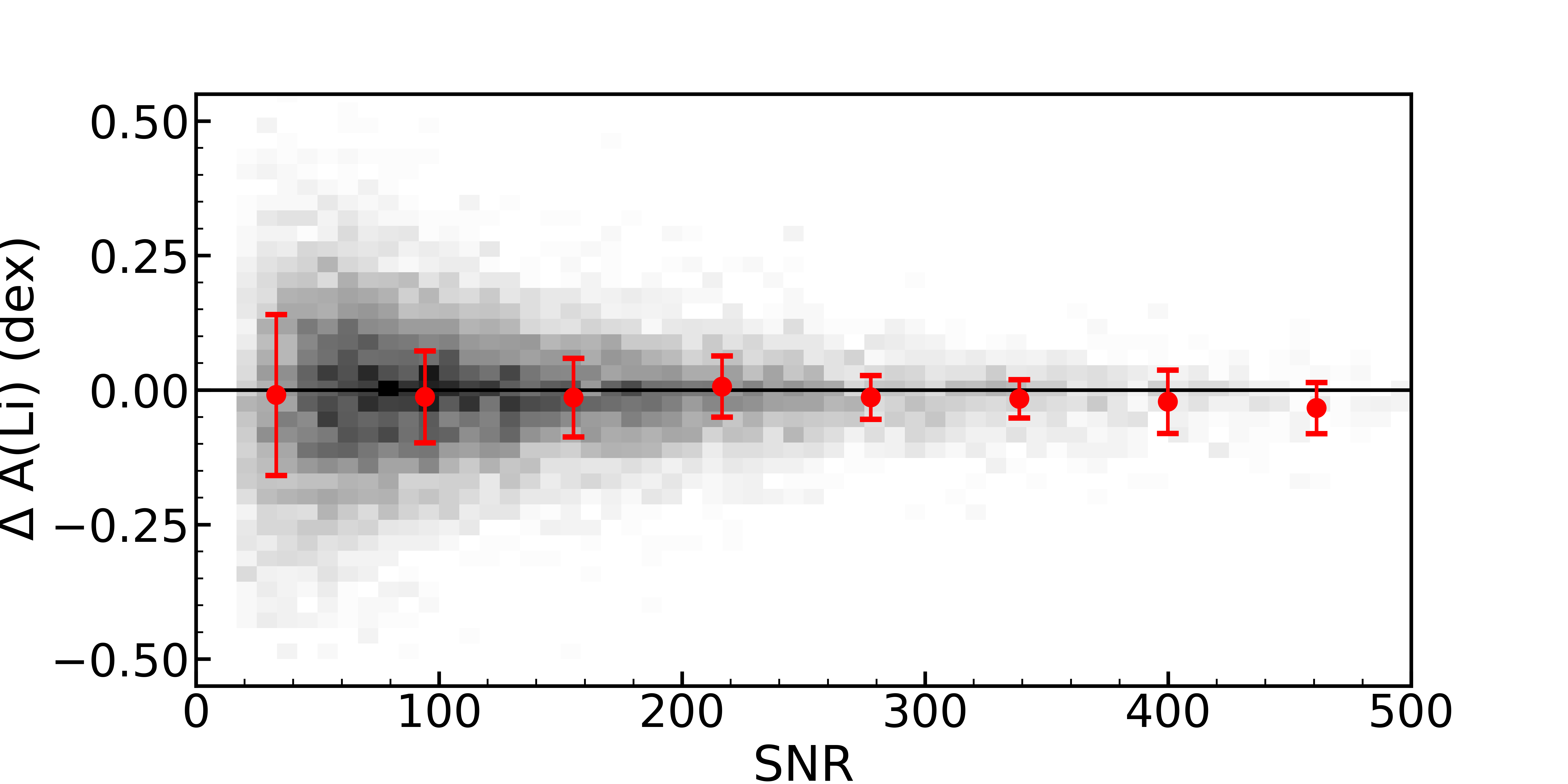}{0.5\textwidth}{}
	\caption{The random errors of \ALi\ as a function of S/N. The red dots and error bars are the average values and the standard deviations of each bin.}
	\label{fig:RandErr}
\end{figure}

Firstly, We make comparisons between repeated observations that have similar spectral quality (the difference in S/N should lower than 20\%).
Figure.~\ref{fig:RandErr} exhibits the difference of \ALi\ as a function of S/N, which suggests that the measurement precision of \ALi\ is greatly influenced by the quality of observed spectra. 
For spectra with S/N $=$ 20, a typical precision of $\sim$ 0.12\,dex is found. 
However, the precision of \ALi\ improves as the S/N of the spectra increases, as expected, and the error can be lower than 0.1\,dex for spectra with better spectral qualities. 

\begin{figure}[!htbp]
	\centering
	\fig{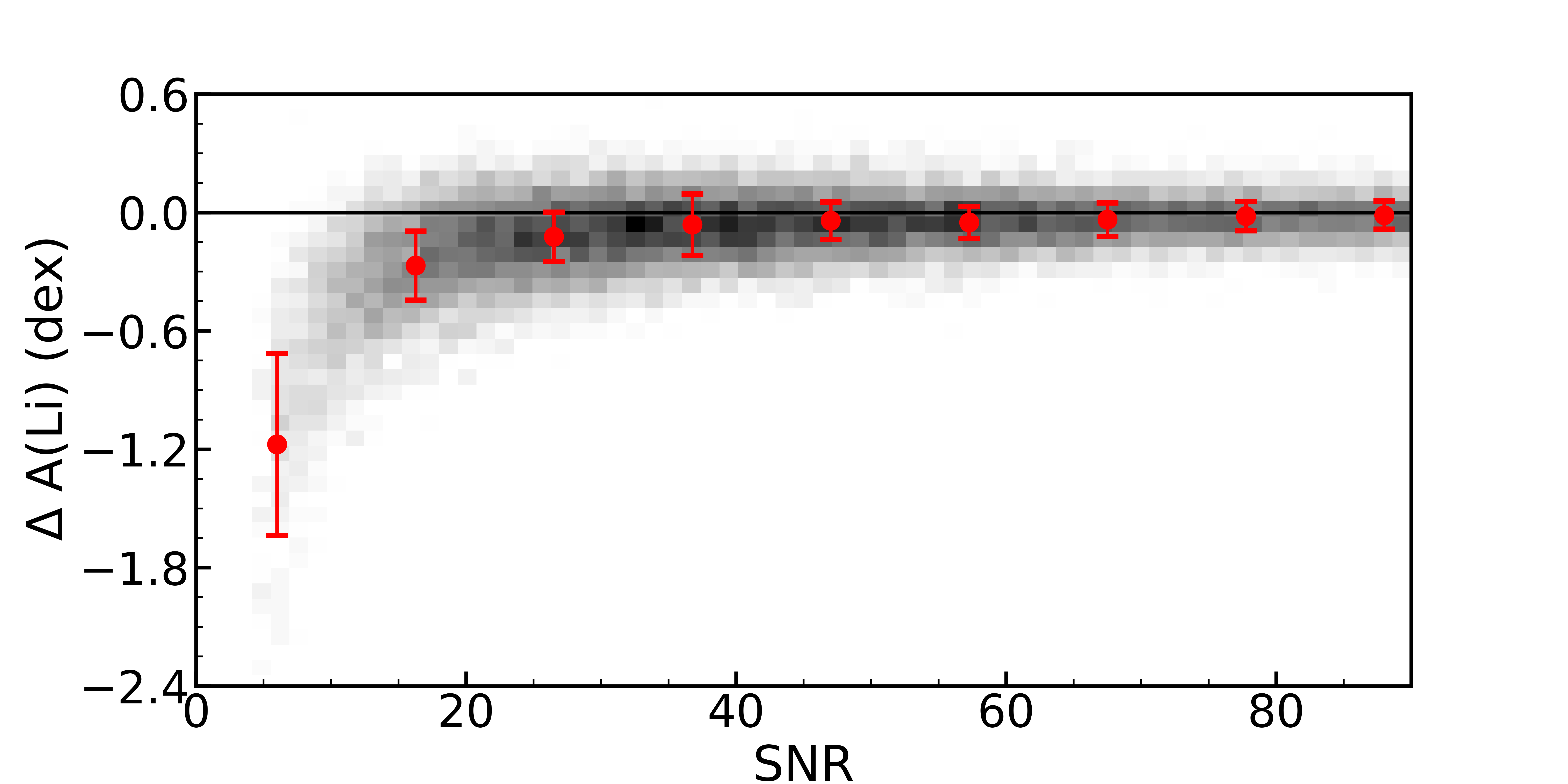}{0.5\textwidth}{}
	\caption{The systematic errors of \ALi\ as a function of S/N. The red dots and error bars are the average values and the standard deviations of each bin.}
	\label{fig:SysErr}
\end{figure}

Secondly, we examine the systematic error that arises from different spectral qualities.
For this purpose, we select spectra of repeated observations for the same stars with a significant difference in spectral quality (the difference in S/N should greater than 40).
Figure.~\ref{fig:SysErr} shows the difference of \ALi\ derived from spectra of repeated observations as a function of the lower S/N. 
The distribution of $\Delta \text{A}(\mathrm{Li})$ is asymmetric, and there is a systematic underestimation for lower resolution samples, which is significantly reduced when S/N reaches 20,
At this point, a typical precision of $\Delta \text{A}(\mathrm{Li})=-0.15$\,dex can be observed.
For spectra with higher S/N, the systematic error between repeated observations becomes smaller.
Therefore, we emphasize that \ALi\ derived from spectra with S/N $>$ 20 are more recommended in order to ensure the reliable statistical analysis.

\section{The Catalog of Lithium Abundance} \label{sec:cat}

The final catalog consists of \numCatalogSpect\ spectra corresponding to \numCatalogStarAll\ unique stars in LAMOST MRS DR9. 
In Table~\ref{Tab:example}, we show an example of our lithium abundance results.
The definition of important attributes is presented in Table~\ref{Tab:colname}.
The spectral identification information includes the designation, the obsid and the source\_id, which represents the LAMOST target ID, the LAMOST unique spectra ID and the \textit{Gaia} DR3 \citep{GaiaCollaboration2022GaiaDataRelease} source identifier.
The max\_diff stands for the largest difference of deduced \ALi\ between the multiple exposures of the same star while the std\_diff means the standard deviation of these measurements, the flag is a ratio between the depth of \ion{Li}{1} 6,707.8\,\AA\ line and the residual fluctuations, the lim suggests whether the upper limit is reached or not. 

\begin{deluxetable}{lll}
	\centering
	\tablecaption{The description of the catalog of lithium abundance.}
	\tabletypesize{\footnotesize}
	\label{Tab:colname}
	\tablehead{\colhead{Column} & \colhead{Unit} & Description }
	\startdata
		desig      &      & the LAMOST target ID                                                      \\
		obsid      &      & the LAMOST unique spectra ID                                              \\
		source\_id &      & source identifier from Gaia DR3                                       \\
		RA         & hms  & right ascension from LAMOST input catalog                             \\
		DEC        & dms  & declination from LAMOST input catalog                                 \\
		S/N        &      & r-band signal-to-noise ratio                                          \\
		RV         & km/s & radial velocity                                                       \\
		T\_eff     & K    & effective temperature                                                  \\
		logg       & dex  & surface gravity                                                       \\
		{[Fe/H]}   & dex  & metallicity                                                            \\
		A(Li)      & dex  & lithium abundance                                                      \\
		max\_diff  & dex  & the maximum of $\Delta \text{A}(\mathrm{Li})$                          \\
		std\_diff  & dex  & the dispresion of $\Delta \text{A}(\mathrm{Li})$         		       \\
		flag       &      & \ion{Li}{1} strength flag                                               \\
		lim        &      & upper limit sign 														\\
		... & ... & ...
	\enddata
\end{deluxetable}

\begin{figure}[!htbp]	
	\centering
	\fig{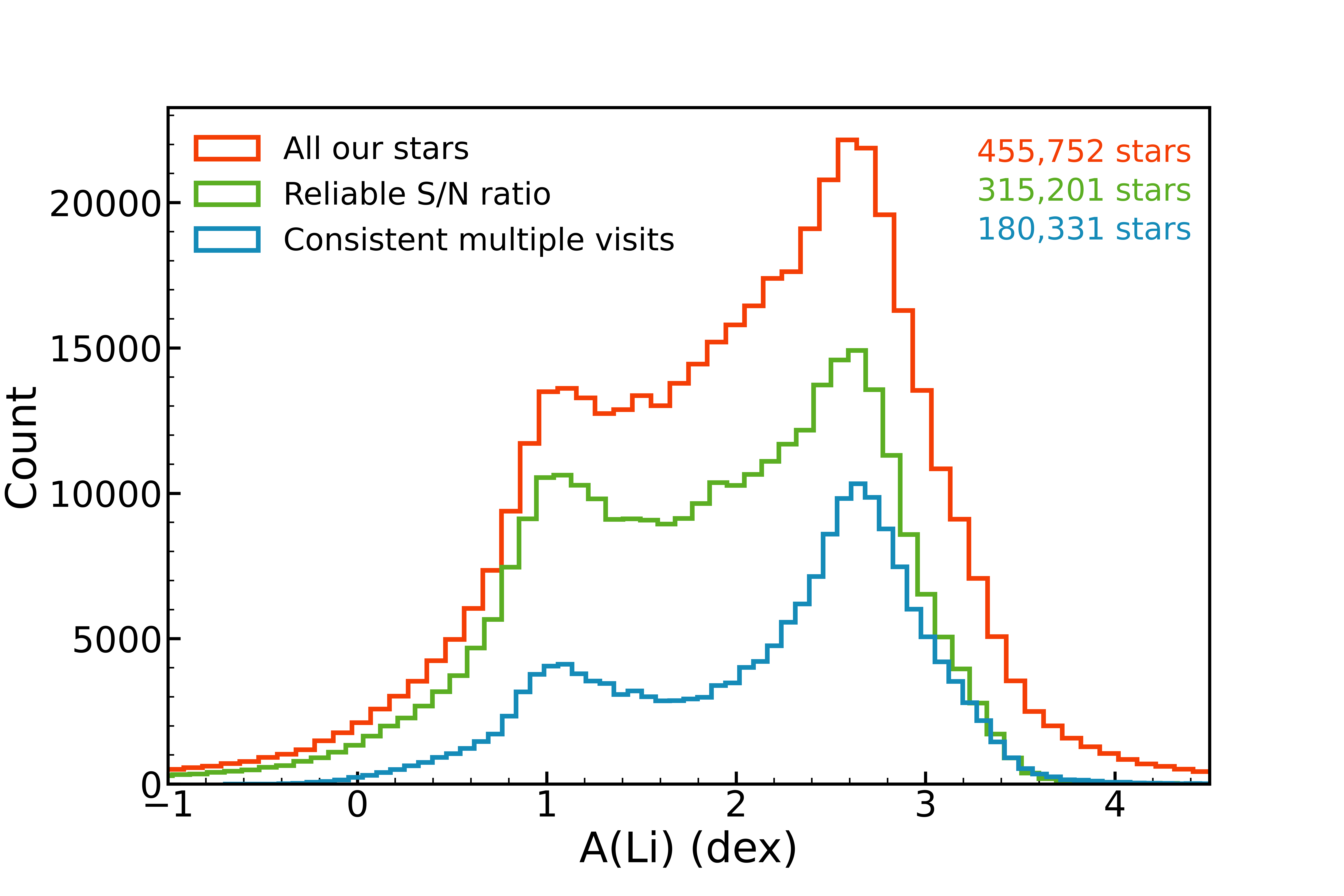}{0.42\textwidth}{}
	\caption{The distribution of \ALi\ for our result. 
	The whole sample set is plotted in red. Samples with reliable observations are plotted in green. The blue distribution shows the samples which have consistent multiple measurements. The number of stars in each distribution are listed respectively.}
	\label{fig:HistALi}
\end{figure}

Figure~\ref{fig:HistALi} illustrates the distribution of \ALi\ for these stars.
In order to better demonstrate our measurements, we apply more restrictions to our samples.
We notice that when we only use the stars with S/N $>$ 20 (the green distribution), the number of stars with \ALi\ $>$ 3.6\,dex are significantly reduced, which suggests a potential effect of noise on the lithium measurements in low S/N spectra.
Next, we further restrict our sample following the criteria below, which is represented by the blue distribution in Figure~\ref{fig:HistALi}.

\begin{enumerate}
	\item{} The upper limits are excluded. 
	\item{} For the multiple exposures of each spectrum, the maximum difference of \ALi\ should less than 1.5\,dex.
	\item{} The dispersion of \ALi\ derived from multiple exposures should less than 0.5\,dex.
\end{enumerate}

As can be seen from Figure~\ref{fig:MRSspec}, the \ALi\ measurement can be very difficult when the \ion{Li}{1} lines is weak, which leads to a decrease in detection efficiency.
As a result, the number of stars with relatively low \ALi\ are significantly reduced.

\begin{figure*}[!htbp]
	\centering
    \gridline{
		\fig{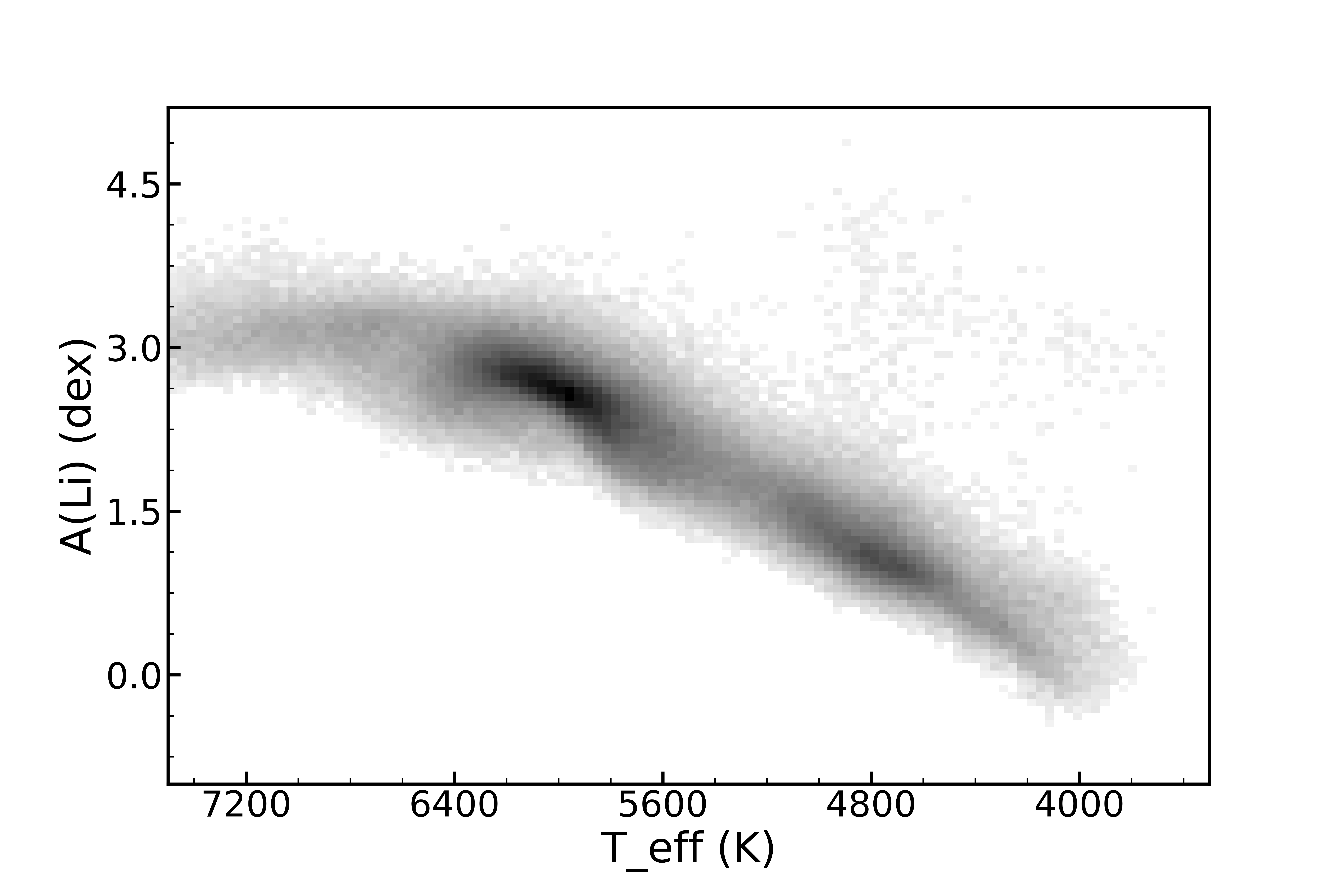}{0.35\textwidth}{}
		\fig{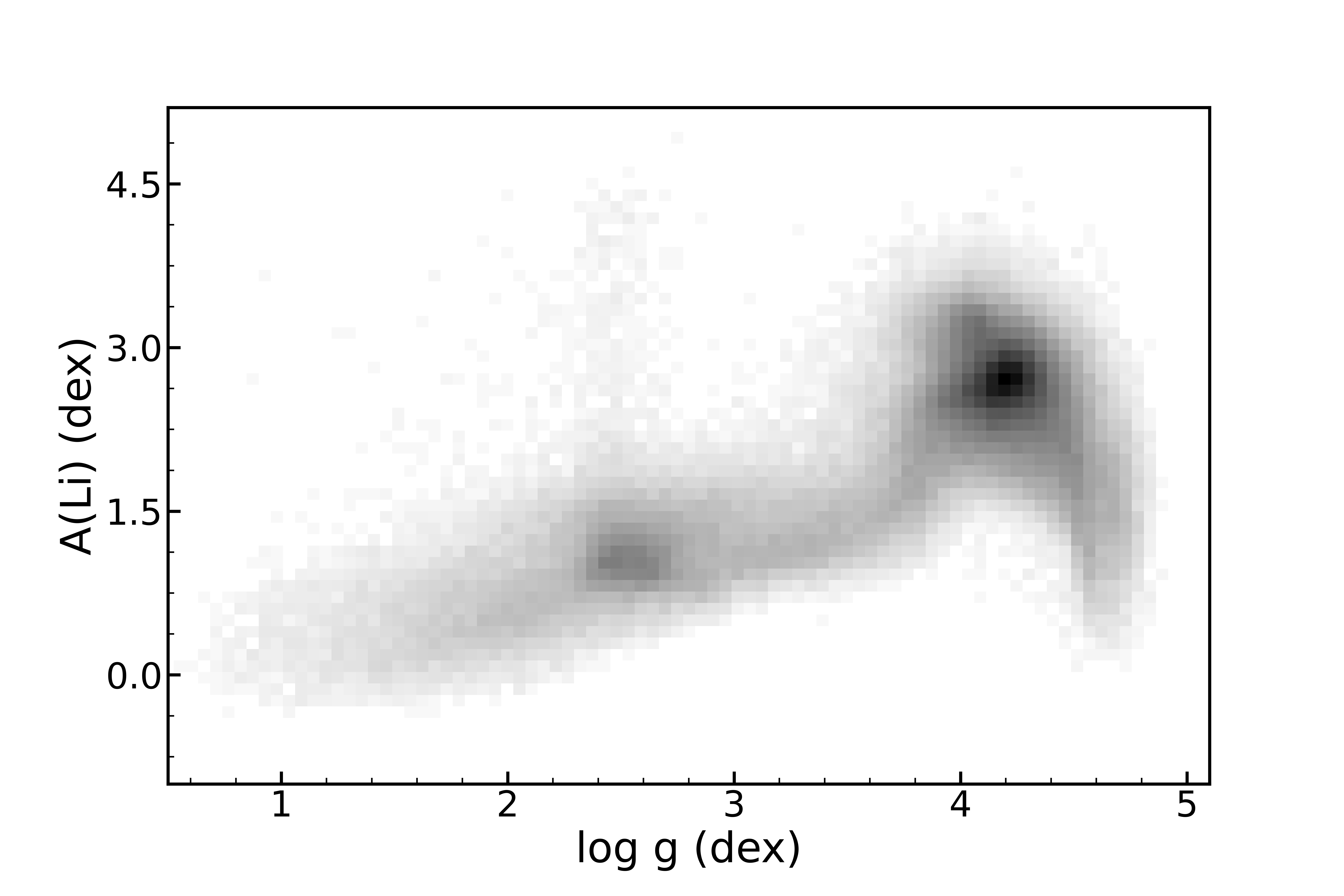}{0.35\textwidth}{}
		\fig{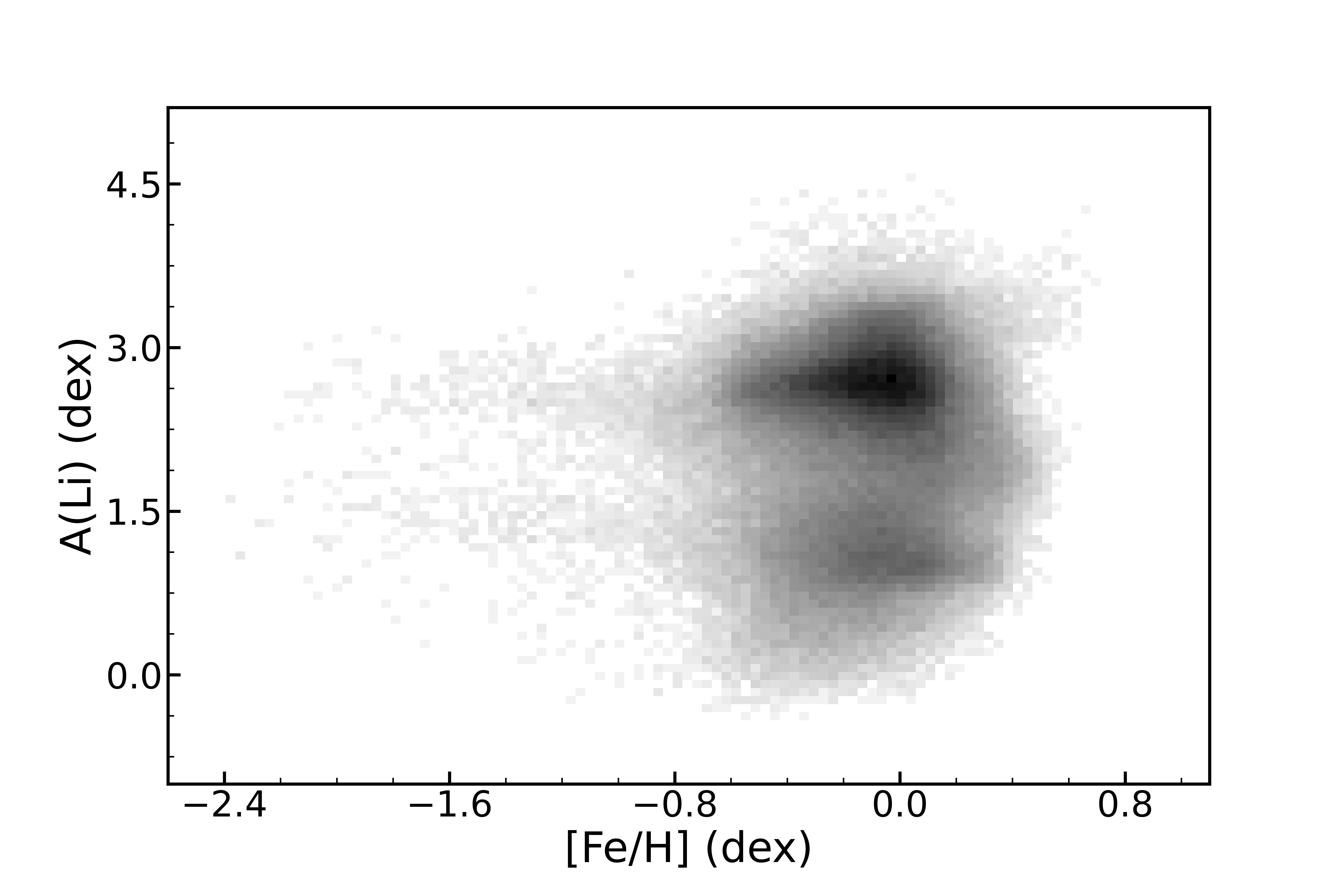}{0.35\textwidth}{}} \vspace{-3mm}
	\caption{The behavior of \ALi\ as functions of \teff, \logg, and \feh. Only stars with S/N $>$ 20 (green distribution in Figure~\ref{fig:HistALi}) are represented.}
	\label{fig:paravsALi}
\end{figure*}

In addition, both distributions in Figure~\ref{fig:HistALi} show two distinct components: one with \ALi\ $\sim$ 2.7\,dex; the other one primarily located at $\sim$ 1.1\,dex.
In order to further investigate these components, we present the distributions of lithium in different stellar parameters in Figure.~\ref{fig:paravsALi}, similar patterns are found in previous studies \citep[e.g.,][]{Gao2021LithiumAbundancesLarge, Buder2021GALAHSurveyThirda}.

In the left panel of Figure.~\ref{fig:paravsALi}, the \teff-\ALi\ relationship is presented, which reveals a decline in lithium abundance as temperature decreases.
We can clearly distinguish that the sample is divided into two separate components, first one is around \ALi\ $\sim$ 2.7\,dex which is mainly hot stars that preserved massive lithium.
While the second component is mainly consisted of cooler stars, suggests that most of them have experienced significant lithium depletion.

The \logg-\ALi\ panel shows that the component of \ALi\ $\sim$ 2.7\,dex is mainly unevolved stars, near 6,000\,K, and they have apparently experienced a mild lithium depletion of an order 0.6\,dex compared to the meteoritic value of $\sim$ 3.3\,dex.
The component with lower \ALi\ is generally giants, most of them have experienced lithium depletion, while few of them, exhibit comparatively high \ALi\ levels, they are so-called Li-rich giants.  
From the literatures \citep{Brown1989SearchLithiumrichGianta, Kumar2011OriginLithiumEnrichment, ruchti2011}, the classical definition of Li-rich should be \ALi\ $>$ 1.5\,dex.
Additionally, \citet{Lagarde2010LiSurveyGiant} suggested that after subgiant lithium dilution is complete (after RGB), stars become cooler than 4,800\,K.
Hence, we use the criterion of \ALi\ $>$ 1.5\,dex, \logg\ $<$ 3.5\,dex and \teff\ $<$ 4,800\,K to identify Li-rich giants from the restricted samples discussed above.
Following the above definition, we eventually identify 976 Li-rich giants, accounting for $\sim 2.6\%$ of the sample stars, consistent with previous studies of $0.5\% \sim 2.2\%$ \citep[e.g.,][]{martell2013a, kirby2016, casey2016a, smiljanic2018, Gao2019LithiumrichGiantsLAMOST, charbonnel2020, yan2020a, martell2021}. 
This is a far higher fraction than what was found for metal-poor stars by \citet{kirby2016} of $\sim$ 0.3\%.
It needs to be pointed out that the \ALi\ threshold for which giants should be properly considered, as Li-rich depends on sensitively on the turnoff \teff\ those stars evolved from, perhaps on their \feh\ and other factors, which means the ratio can greatly depend on the adopted definition of Li-rich giants.
Furthermore, we observed 174 giants with lithium abundances exceed the primordial lithium abundance of 3.3\,dex, which are classified as super Li-rich giants, comprising $\sim 18\%$ of all our Li-rich giants. 
The lithium content in these super Li-rich giants is surprisingly high, which makes them valuable samples in further studies. 
Perhaps most of them undergo a phase of lithium production followed by its destruction, so only a small fraction of giants is observed to be Li-rich at any given time.
Notably, we also discover 335 super Li-rich unevolved candidates that have lithium abundances higher than 3.8\,dex, which significantly surpasses the meteoritic value of 3.3\,dex. 
These dwarfs, which have extremely shallow surface convection zones, examine extraordinarily high lithium abundance levels that challenge current models.
More specifically, the mechanisms such as accretion of rocky planetesimals and upwards diffusion (for example, radiative acceleration), which could bring lithium to the surface from below and create super-meteoritic values \citep{Richer1993DiffusionLithiumBeryllium}, could possibly explain the unusual enrichment.

The \feh-\ALi\ relationship is also an important indicator of lithium evolution, for that purpose, we trace \ALi\ as a function of metallicity in the right panel of Figure.~\ref{fig:paravsALi}.
We clearly see that dwarfs with $-2.4$ $<$ \feh\ $<$ $-1.6$\,dex show a constant lithium abundance, which is very close to the plateau of SBBN abundance value.
Thus, we visually checked the quality of the fit between the observed and synthetic spectrum and find six metal-poor stars which also have lithium abundance exceeds the meteoritic abundance value. 
One of these stars (J1314$+$3741) has been analyzed by \citet{li2018a} with high dispersion spectrum \citep[HDS,][]{noguchi2002} observed with the \textit{Subaru Telescope} having $\text{A}(\mathrm{Li})_{\text{LTE}}=$ 3.48\,dex.
While our \ALi\ measurement of J1314$+$3741 is 3.56\,dex using the same stellar parameters from \citet{li2018a}.
A time variation of the radial velocity is detected for this object, hence, the interpretation of it's high \ALi\ is considered to be the accretion of matter from a highly evolved companion star or an extra mixing caused by merging events with planets or other small-mass objects.
Considering our results have a systematic overestimation of $\sim$ 0.1\,dex which is discussed in Section~\ref{sec:valid}, we think it is a consistent result.

It should be emphasized that our catalog is the first time that carried out such a large statistical and homogeneous sample size of lithium abundances from LAMOST DR9 through a very wide stellar parameter range.

\section{Conclusion} \label{sec:con}

In this study, we employ a template-matching method to determine the lithium abundance for the LAMOST MRS by fitting the \ion{Li}{1} resonance line.
We finally derive the lithium abundance of \numCatalogSpect\ spectra for \numCatalogStarAll\ stars from the LAMOST MRS DR9.
The comparisons between our measurements and the high-resolution references such as GALAH and \textit{Gaia}-ESO show good agreements with a deviation of $\sigma A(\mathrm{Li}) \sim$ \sigGALAH.
To evaluate the internal precision of our method, we perform an error analysis based on stars that have multiple observations from LAMOST MRS.
The random error of our \ALi\ measurement is sensitive to the observational quality, meanwhile, a systematic overestimation was observed for spectra with low S/N.
However, for spectra with S/N higher than 20, the internal error becomes smaller, and the typical precision can be better than 0.1\,dex for these stars.
The catalog of lithium abundance is presented and $\sim$ 900 Li-rich giants are found during a provisional analysis of our results.
A larger number giants may be Li-rich if lithium depletion during the main sequence is also taken into account.
Based on these Li-rich samples, multiple stars with extraordinary lithium enrichment are identified. 
The chemical and kinematic properties of these results are worth further exploration, which will be helpful to unveil the mist of lithium evolution.

We supply a LAMOST MRS DR9 catalog with massive consistent of lithium abundance measurements, and the template-matching method constructed in this work will be applicable to further medium-resolution observations like the ongoing LAMOST MRS DR10. 

\begin{acknowledgments}
We thank the anonymous referee for the careful reading of the manuscript and useful comments.
Our research is supported by the National Natural Science Foundation of China
under grant Nos. 12090040, 12090044, 11833006, 12373036, 12022304, 11973052, 12103063 and the National Key R\&D Program of China No.2019YFA0405502. 
We acknowledge the science research grants from the China Manned Space Project with NO.CMS-CSST-2021-B05.
This work is supported by Chinese Academy of Sciences President's International Fellowship Initiative. Grant No. 2020VMA0033.
H.-L.Y. acknowledges the supports from Youth Innovation Promotion Association, Chinese Academy of Sciences.
Guoshoujing Telescope (the Large Sky Area Multi-Object Fiber Spectroscopic Telescope LAMOST) is a National Major Scientific Project built by the Chinese Academy of Sciences.
Funding for the project has been provided by the National Development and Reform Commission.
LAMOST is operated and managed by the National Astronomical Observatories, Chinese Academy of Sciences.

\end{acknowledgments}

\appendix
\section{Appendix Table} \label{sect:appendixTable}

\begin{longrotatetable}
\label{Tab:example}
\begin{deluxetable*}{rrrrrrrrrrrrrrr}
\tablewidth{800pt}
\tablecaption{Lithium abundance of randomly selected stars.}
\centering
\tablehead{\colhead{desig} & \colhead{obsid} & \colhead{source\_id} & \colhead{RA} & \colhead{DEC} & \colhead{S/N} & \colhead{RV} & \colhead{T\_eff} & \colhead{log g} & \colhead{[Fe/H]} & \colhead{A(Li)} & \colhead{max\_diff} & \colhead{std\_diff} & \colhead{flag} & \colhead{lim}\\ 
\colhead{} & \colhead{} & \colhead{} & \colhead{hms (J2000)} & \colhead{dms (J2000)} & \colhead{} & \colhead{(km$\cdot s^{-1}$)} & \colhead{(K)} & \colhead{(dex)}  & \colhead{(dex)} & \colhead{(dex)} & \colhead{(dex)} & \colhead{(dex)} & \colhead{} & \colhead{}} 
    \startdata
	J$023628.56+340501.3$ & $ 786603015$ & $ 134865653886023552 $ & $ 02:36:28.56$ & $ +34:05:01.3$ & $ 50        $ & $ -67      $ & $ 4923       $ & $ 2.57       $ & $ 0.00        $ & $ 1.3         $ & $ 0.3             $ & $ 0.1              $ & $ 1          $ & $ 0 $ \\
	J$112411.19+544301.3$ & $ 726408210$ & $ 843172689467150336 $ & $ 11:24:11.19$ & $ +54:43:01.4$ & $ 82        $ & $ -52      $ & $ 5738       $ & $ 4.44       $ & $ 0.06        $ & $ 2.1         $ & $ 0.1             $ & $ 0.1              $ & $ 2          $ & $ 0 $ \\
	J$081045.62+203639.0$ & $ 609816029$ & $ 675898492272015488 $ & $ 08:10:45.62$ & $ +20:36:39.0$ & $ 82        $ & $ 27       $ & $ 4662       $ & $ 2.43       $ & $ 0.08        $ & $ 1.9         $ & $ 0.0             $ & $ 0.0              $ & $ 10         $ & $ 0 $ \\
	J$191337.83+470954.9$ & $ 591905138$ & $ 2130795024295529728$ & $ 19:13:37.84$ & $ +47:09:55.0$ & $ 82        $ & $ -78      $ & $ 4444       $ & $ 2.31       $ & $ 0.06        $ & $ 0.7         $ & $ 0.1             $ & $ 0.1              $ & $ 1          $ & $ 0 $ \\
	J$005419.69+101508.0$ & $ 845514222$ & $ 2582346548394869504$ & $ 00:54:19.70$ & $ +10:15:08.0$ & $ 59        $ & $ -19      $ & $ 5781       $ & $ 4.45       $ & $ -0.13       $ & $ 2.1         $ & $ 0.5             $ & $ 0.2              $ & $ 2          $ & $ 0 $ \\
	J$192306.05+480524.4$ & $ 824903227$ & $ 2129404726202499456$ & $ 19:23:06.05$ & $ +48:05:24.5$ & $ 76        $ & $ -14      $ & $ 4619       $ & $ 2.48       $ & $ 0.09        $ & $ 0.9         $ & $ 0.1             $ & $ 0.0              $ & $ 1          $ & $ 0 $ \\
	J$170735.29+324454.1$ & $ 809911091$ & $ 1334528153701029760$ & $ 17:07:35.29$ & $ +32:44:54.1$ & $ 82        $ & $ 1        $ & $ 4419       $ & $ 2.19       $ & $ -0.10       $ & $ 0.6         $ & $ 0.1             $ & $ 0.0              $ & $ 1          $ & $ 0 $ \\
	J$093743.49+502035.3$ & $ 766701059$ & $ 826265056786653568 $ & $ 09:37:43.50$ & $ +50:20:35.4$ & $ 82        $ & $ -30      $ & $ 5466       $ & $ 3.88       $ & $ -0.07       $ & $ 2.2         $ & $ 0.0             $ & $ 0.0              $ & $ 6          $ & $ 0 $ \\
	J$034609.62+231811.0$ & $ 785405177$ & $ 64970971016423040  $ & $ 03:46:09.63$ & $ +23:18:11.0$ & $ 64        $ & $ 15       $ & $ 5345       $ & $ 3.85       $ & $ -0.10       $ & $ 1.6         $ & $ 0.2             $ & $ 0.1              $ & $ 2          $ & $ 0 $ \\
	J$123342.30+400618.7$ & $ 636311187$ & $ 1533614696916850816$ & $ 12:33:42.30$ & $ +40:06:18.7$ & $ 82        $ & $ -40      $ & $ 4750       $ & $ 2.43       $ & $ -0.44       $ & $ 1.0         $ & $ 0.3             $ & $ 0.1              $ & $ 1          $ & $ 0 $ \\
	J$082631.28+110300.3$ & $ 817309148$ & $ 601097200109362560 $ & $ 08:26:31.29$ & $ +11:03:00.4$ & $ 70        $ & $ 74       $ & $ 4811       $ & $ 2.94       $ & $ -0.08       $ & $ 1.2         $ & $ 0.1             $ & $ 0.1              $ & $ 1          $ & $ 0 $ \\
	J$150220.11+222559.4$ & $ 822603243$ & $ 1261982818880193408$ & $ 15:02:20.11$ & $ +22:25:59.5$ & $ 78        $ & $ -7       $ & $ 5208       $ & $ 4.53       $ & $ 0.09        $ & $ 1.5         $ & $ 0.1             $ & $ 0.0              $ & $ 1          $ & $ 0 $ \\
	J$085208.51+115540.3$ & $ 808208093$ & $ 604943639676920448 $ & $ 08:52:08.52$ & $ +11:55:40.4$ & $ 58        $ & $ 5        $ & $ 5543       $ & $ 3.75       $ & $ -0.09       $ & $ 2.3         $ & $ 0.1             $ & $ 0.0              $ & $ 5          $ & $ 0 $ \\
	J$163924.70+282835.7$ & $ 747207220$ & $ 1310869893286754048$ & $ 16:39:24.71$ & $ +28:28:35.7$ & $ 81        $ & $ -58      $ & $ 4685       $ & $ 2.39       $ & $ -0.20       $ & $ 0.9         $ & $ 0.1             $ & $ 0.0              $ & $ 1          $ & $ 0 $ \\
	J$050156.33+233646.2$ & $ 685213076$ & $ 3418630518340608384$ & $ 05:01:56.34$ & $ +23:36:46.2$ & $ 63        $ & $ 32       $ & $ 6134       $ & $ 4.29       $ & $ -0.07       $ & $ 2.8         $ & $ 0.2             $ & $ 0.1              $ & $ 4          $ & $ 0 $ \\
	J$135134.75+393959.7$ & $ 636612026$ & $ 1496983917281074944$ & $ 13:51:34.75$ & $ +39:39:59.8$ & $ 82        $ & $ -14      $ & $ 5555       $ & $ 4.46       $ & $ 0.08        $ & $ 1.6         $ & $ 0.1             $ & $ 0.0              $ & $ 1          $ & $ 0 $ \\
	J$190102.29+415821.6$ & $ 591703223$ & $ 2104170968704244608$ & $ 19:01:02.30$ & $ +41:58:21.6$ & $ 82        $ & $ -38      $ & $ 4556       $ & $ 2.42       $ & $ 0.24        $ & $ 1.0         $ & $ 0.1             $ & $ 0.1              $ & $ 2          $ & $ 0 $ \\
	J$074904.91+285751.7$ & $ 715707199$ & $ 875795371262406272 $ & $ 07:49:04.92$ & $ +28:57:51.7$ & $ 82        $ & $ -13      $ & $ 6088       $ & $ 3.92       $ & $ -0.42       $ & $ 2.4         $ & $ 0.2             $ & $ 0.1              $ & $ 2          $ & $ 0 $ \\
	... &  ... & ... & ... & ... & ... & ... & ... & ... & ... & ... & ... & ... & ... & ... \\
	\enddata
	 \tablecomments{The attribute column is not fully presented here. The complete catalog is accessible in its entirety in a machine-readable form.}
\end{deluxetable*}
\end{longrotatetable}

\bibliography{sample631}{}
\bibliographystyle{aasjournal}

\end{document}